\documentclass[a4paper,twocolumn,showpacs,superscriptaddress,floatfix]{revtex4-2}
\usepackage{graphicx}
\usepackage{epsf}
\usepackage{epsfig}
\usepackage{amssymb,amsmath}
\usepackage[usenames]{color}
\usepackage{amssymb}
\usepackage{times}
\usepackage{bbold}

\newcommand{\w}[1]{\mathbf{#1}}
\newcommand{\be}{\begin{equation}}
\newcommand{\ee}{\end{equation}}
\newcommand{\bea}{\begin{eqnarray}}
\newcommand{\eea}{\end{eqnarray}}
\newcommand{\nn}{\nonumber}

\font\tenscr=rsfs10 scaled1100
\font\sevenscr=rsfs7 
\font\fivescr=rsfs5 
\skewchar\tenscr='177
\skewchar\sevenscr='177
\skewchar\fivescr='177
\newfam\scrfam
\textfont\scrfam=\tenscr
\scriptfont\scrfam=\sevenscr
\scriptscriptfont\scrfam=\fivescr

\begin{document}

\title[Airy-function approach to binary black hole merger waveforms: The fold-caustic model]
{Airy-function approach to binary black hole merger waveforms: The fold-caustic diffraction model}


\author{Jos\'e Luis Jaramillo}
\address{Institut de Math\'ematiques de Bourgogne (IMB), UMR 5584,
  CNRS, Universit\'e de Bourgogne Franche-Comt\'e, F-21000 Dijon,
  France}

\author{Badri Krishnan}
\address{Institute for Mathematics, Astrophysics and Particle Physics,
  Radboud University, Heyendaalseweg 135, 6525 AJ Nijmegen, The
  Netherlands}
\address{Max Planck Institute for Gravitational Physics (Albert
  Einstein Institute), Callinstrasse 38, D-30167 Hannover, Germany}

\vspace{10pt}

\begin{abstract}
  
  From numerical simulations of the Einstein equations, and also from
  gravitational wave observations, the gravitational wave signal from
  a binary black hole merger is seen to be simple and to possess
  certain universal features.  The simplicity is somewhat surprising
  given that non-linearities of general relativity are thought to play
  an important role at the merger.  The universal features include an
  increasing amplitude as we approach the merger, where transition
  from an oscillatory to a damped regime occurs in a pattern
  apparently oblivious to the initial conditions. We
  propose an Airy-function pattern
  to model the binary black hole (BBH) merger waveform, focusing on
  accounting for its simplicity and universality.  We postulate that
  the relevant universal features are controlled
  by a physical mechanism involving:
  i) a caustic phenomenon in a basic `geometric optics'
  approximation and, ii) a diffraction over the caustic
  regularizing its divergence.  Universality of caustics and their
  diffraction patterns account for the observed universal features, as in
  optical phenomena such as rainbows.
  This postulate, if true, allows us to borrow mathematical techniques
  from Singularity (Catastrophe) Theory,  in particular  Arnol'd-Thom's theorem,
  and to understand binary mergers in terms of fold caustics.  The diffraction pattern
  corresponding to the fold-caustic is given in terms of the Airy
  function, which leads (under a `uniform approximation') to
  the waveform model written in terms of a parameterized Airy
  function.  The post-merger phase does not share the same features of
  simplicity and universality, and must be added separately.
  Nevertheless, our proposal allows a smooth matching of the inspiral
  and post-merger signals by using the known asymptotics of the Airy
  function.

\end{abstract}

\pacs{}

\maketitle


\section{Introduction: the simplicity of compact binary merger waveforms}

\subsection{Compact binary merger waveforms: a status overview}
The gravitational wave signal emitted during the merger of two black
holes is now calculated routinely by numerical simulations, at least
for moderate mass ratios.  These numerical simulations solve the
Einstein equations with appropriate initial data.  The first numerical
simulations of binary black hole (BBH) mergers were successfully
carried out in 2005
\cite{Pretorius:2005gq,Campanelli:2005dd,Baker:2005vv}.  Extensive
public catalogs of numerically generated BBH waveform are available
\cite{Boyle:2019kee,Healy:2020vre,Jani:2016wkt}.  On the observational
front, signals from binary black hole mergers were first detected by
the LIGO and Virgo detectors in 2015 \cite{Abbott:2016apu}.  Since the
first detection, the LIGO and Virgo detectors have carried out
extensive observational runs with increasing sensitivity, and close to
a hundred merger events have now been observed; see
e.g. \cite{LIGOScientific:2021djp,Nitz:2021zwj} for the most recent
event catalogs.  The Japanese KAGRA detector has also started to
participate in the observing runs \cite{LIGOScientific:2022myk}.  The
vast majority of these events are binary black hole mergers, though
two binary neutron star mergers and two neutron star - black hole
mergers have been detected.  The results of the numerical simulations
are so far seen to be fully consistent with gravitational wave
observations by the LIGO and Virgo detectors (see
e.g. \cite{LIGOScientific:2016kms}).  Furthermore numerical relativity
results are used in different ways to calibrate analytical waveform
models for binary mergers, which are then used to filter gravitational
wave data.

These observations and simulations confirm the presence of the three
well known regimes of binary black hole mergers, namely the
inspiral, merger and ringdown.  The early part of the signal is the
inspiral, where the two black holes orbit each other and the orbit
gradually shrinks due to energy loss by emission of gravitational
radiation.  The signal here is quasi-periodic with an increasing
amplitude and frequency and can be calculated to a good approximation
by post-Newtonian techniques (see e.g. \cite{Bla06,Porto:2016pyg}).  The late part of
the signal is the ringdown, where the final black hole has formed and
is approaching its final equilibrium state. Here the signal is a
superposition of exponentially damped oscillations, and the
frequencies and damping times can be computed within the framework of
black hole perturbation theory (see e.g. \cite{Kokkotas:1999bd}).
Thus far, no consistent approximation scheme can deal with the merger
regime and one generally relies on the aforementioned numerical
simulations (or analytical models calibrated by these simulations).

A suitable approximation scheme specifically devoted for the merger
would be highly desirable, especially in the forthcoming era of third
generation gravitational wave detectors. The accuracy requirements on
waveforms will become significantly more stringent over the next
decade as detectors become more sensitive and can discern finer
features of the waveforms.  Using less accurate waveforms will likely
lead one to draw incorrect conclusions.  A clear example is in tests
of general relativity and searches for new physics.  This problem
extends also to binary neutron star mergers.  Constraints on the
neutron star equation of state rely on measurements of neutron star
tidal deformability.  These appear as higher order terms in the
post-Newtonian approximation, and contribute in the late inspiral
regime \cite{Flanagan:2007ix,Damour:2009vw}. Thus, an incorrect merger
signal could lead to incorrect measurements of the tidal deformability
and eventually to incorrect constraints on the equation of state.  Yet
another example is the memory effect.  This is a non-linear effect in
general relativity which leads to a permanent displacement in the arms
of an interferometric detector
\cite{Christodoulou:1991cr,Frauendiener_1992,Thorne:1992sdb,Blanchet:1992br,Hubner:2019sly,Lasky:2016knh,Favata:2009ii,Khera:2020mcz}.
The contributions to the memory are largest in the merger and thus,
more accurate merger models would aid in the quest for direct
observations of the memory effect.

\subsection{Simplicity, universality and `effective linearity' of BBH
  merger waveforms}

\subsubsection{Simplicity and Universality of merger waveforms}

Before the first successful numerical simulations of the merger in
2005, it was thought plausible that the waveform could be very complex
as we transition from the inspiral to the merger; see e.g. the
illustration due to Kip Thorne shown in Fig. 1 of
\cite{Schutz:2004uj}.  This figure might suggest that the
non-linearities of general relativity could lead to the presence of
complicated modulations in the merger signal which might depend
sensitively on initial conditions of the binary.  However, nature
turns out out to be much simpler and no such complexity is observed.
Indeed, this simplicity is partly responsible for the success of the
waveform models mentioned above.  On the other hand, the post-merger
signal is not universal and depends on the compact object
nature of the remnant that is formed by the merger.  Thus, for a
binary black hole merger, the post-merger signal is determined by the
ringing down modes of the remnant black hole. In contrast, for a
binary neutron star merger leading initially to the formation of a
hyper-massive neutron star, the post-merger signal can be much more
complicated.  For this reason as we shall see, the post-merger signal
needs to be treated very differently.

We should note here that even though the waveform might appear simple,
it encodes a non-trivial structure. There is detailed information
hidden in the precise manner in which the amplitude and frequency
evolve, and modeling this accurately is crucial to extracting reliable
information from gravitational wave observations.  The fact that the
gravitational waveform does not have any wild modulations does not
mean that the merger signal does not have important physical
information. Rather, it means that the complexities are to be found in
small deviations from an underlying simple solution, and most of the
difficulty lies in correctly identifying this dominant solution.  This
simplicity has implicitly helped in the development of several
successful complete waveform models which accurately model the
inspiral, merger and ringdown phases (see
e.g. \cite{Buonanno:1998gg,Nagar:2018zoe,Damour:2016bks,Bohe:2016gbl,Taracchini:2013rva,Khan:2019kot,Khan:2018fmp,London:2017bcn,Khan:2015jqa,Husa:2015iqa,Santamaria:2010yb,Ajith:2009bn,Ajith:2007kx}).
Using a combination of analytic and numerical methods, these complete
waveform models have seen continual improvements over the last decade
and especially since the first detections. They are now capable of
modeling increasingly non-trivial physical effects such as precession,
eccentricity, and can now also incorporate higher angular modes.

In contrast with this approach, we shall not seek to improve waveform
models by adding more complexity.  Our strategy will rather be to
simplify to the largest extent possible.  To this end, we identify the
following key (`universal') qualitative features of the gravitational
waveform emitted during a binary black hole merger waveform:
\begin{enumerate}
\item In the inspiral regime, in the context of the post-Newtonian
  approximation, the signal is a ``chirp'' signal with increasing
  amplitude and frequency. Crucially, the post-Newtonian amplitude formally
  diverges as we approach the merger and the approximation breaks
  down. 
\item Across the merger, the qualitative nature of the signal changes
  dramatically (`catastrophically') from an oscillatory behavior in
  the pre-merger regime, to a damped behavior post-merger.  Thus, the
  gravitational wave amplitude increases up to the merger, and
  ``shuts-off'' very soon after the merger.
\item The post-merger signal is a combination of damped sinusoids for
  binary black hole mergers.  This approximation breaks down as we go
  backwards in time towards the merger.  We note that the signal is
  possibly much more complicated for mergers involving neutron stars.
\end{enumerate}
The features of the inspiral and merger waveforms are qualitatively
similar for mergers involving neutron stars, though the post-merger
regime can be dramatically different.  These features and the apparent
simplicity and universality deserve an explanation.  We can also ask:
should the merger waveform in modified gravity theories also display
this simplicity and universality?

\subsubsection{Asymptotic Reasoning and Structural Stability}

Our strategy is based on two tenets: the Simplicity and Universality
of the BBH merger waveforms.  These two principles are
respectively implemented through the notions of `asymptotic reasoning'
and 'structural stability'.

The first principle is, as mentioned above, to simplify the problem as
far as possible to focus on its structural aspects.  Such an approach
is captured in what can be referred to as ``asymptotic reasoning''
\cite{Batte01}, that aims to uncover the qualitative mechanisms and
patterns underlying the observed simplicity~\footnote{The notion of
  `asymptotic reasoning' is discussed by Batterman in~\cite{Batte01}:
  ``this type of reasoning involves, at its heart, a type of
  abstraction ---a means for ignoring or throwing away various
  details.  [...] I call this kind of reasoning `asymptotic reasoning'
  ``.
  Slightly rephrased also in Batterman~\cite{Batte01}: ``scientific
  understanding often requires methods which eliminate detail and, in
  some sense, precision [...]. I call these methods `asymptotic
  methods' and the type(s) of reasoning they involve `asymptotic
  reasoning'.}.  For our purposes, this leads us to disregard effects
such as precession, eccentricity etc. which, though physically
important, do not modify the essential features of the waveform.
These details can be re-introduced at a later stage once the waveform
features mentioned above are understood.  Moreover, from a more
technical perspective, this `asymptotic reasoning' indeed involves an
asymptotic limit in terms of an appropriate parameter, in the spirit
that asymptotics do capture the relevant underlying patterns, that can
(often) be extended beyond the proper asymptotic limit.  In our case,
such an asymptotic treatment will be provided using an analogy with
geometric optics.

The second principle is that of Universality, which refers to the fact
that the qualitative features of the merger waveform do not depend on
the details on the initial configuration. Universality will be
captured and implemented by the notion of structural stability,
i.e. stability under small generic perturbations.  Here we recognize that
mathematical models of natural phenomena are never exact, and thus it
is essential for our models to be robust under small changes of the
initial conditions and also in fact, under small changes of the
dynamical equations.  In particular, under small deformations of
general relativity itself, we should expect the above-mentioned
essential features of BBH waveforms to persist.  Finally, it can be
argued that Simplicity and Universality are not independent
notions. Indeed, a condition for discussing `asymptotic reasoning' is
the realization of some kind of `structural
stability'~\cite{Batte01}. It will be useful though to
methodologically separate these two aspects.

The idea that BBH waveform should be simple and universal is not a new
one.  In fact, even before the results of the numerical simulations
were available, there were several prescient suggestions of such
simplicity.  A notable example is the early Effective-One-Body model
proposed by Damour, Buonanno and collaborators
\cite{Buonanno:2000ef,Buonanno:1998gg}.  Their work was based on two
fundamental ingredients: i) gravitational radiation leads to a
circularization of the orbits of the binary system as it goes through
the inspiral regime in a sequence of Keplerian orbits. ii) This
``adiabatic'' inspiral phase terminates as the system reaches a
last-stable orbit.  This causes a qualitative transition from the
inspiral to a ``plunge'' phase (followed shortly by the merger and
ringdown).  This suggests a Universality of the merger signal and is
reminiscent of the effacement principle studied by Damour
\cite{1987thyg.book..128D}, where the internal structure of the two
binary components does not appear in the signal at leading order.
Apart from this Universality, the Buonanno-Damour proposal suggests
that the merger should be simple as well, which in turns contributes
to the success of this approach.

A second notable support to Simplicity and Universality is the
Close-Limit approximation due to Pullin, Price, Gleiser, Khanna and
collaborators \cite{Price:1994pm,Gleiser:1996yc,Khanna:1999mh}.  It
adds an additional ingredient, a sort of `effective linearity', that
we expand below.

In summary, simplicity partly explains the success of existing signal
models which have been employed in various analyses to date (and are
largely responsible for many of the remarkable achievements summarized
above).  Developing such signal models would have been much more
challenging if the late inspiral were to have additional features
depending sensitively under small variations of physical parameters.

\subsubsection{`Effective linearity' of the BBH merger waveform}

The Close-Limit approximation mentioned above shows
that in the very late inspiral phase, when the two black holes are
very close to each other, the spacetime can be surprisingly well
approximated as the perturbation of a single black hole.  This
approach has been used to estimate the recoil velocity of the final
black hole \cite{Sopuerta:2006wj,Sopuerta:2006et}.  We
mention also the Lazarus project \cite{Baker:2001nu,Baker:2001sf}
which, partly motivated by the success of the Close-Limit
approximation, combined the results of numerical relativity with
perturbation theory.  Again, this approach suggests both universality
and simplicity. 

The surprising success of the close-limit approximation suggests an
``effective linearity'' underlying the essential mechanism behind
binary black hole merger waveforms.  There are other hints as well for
linearity in black hole mergers: i) It has been suggested that the
linear description of the ringdown phase can be extended all the way
back to the merger itself
\cite{Giesler:2019uxc,Isi:2019aib,Okounkova:2020vwu,Capano:2021etf}
(though several open questions remain
\cite{Cotesta:2022pci,Isi:2022mhy}). ii) Other studies find strong
evidence for the presence of linear correlations between geometric
fields in the strong field region and fields in the asymptotic
wave-zone in a BBH merger
\cite{Jaramillo:2011re,Jaramillo:2011rf,Jaramillo:2012rr,Gupta:2018znn,Prasad:2020xgr,Iozzo:2021vnq,Ashtekar:2021wld,Ashtekar:2021kqj}.
These correlations are a dynamical feature difficult to understand
without accepting some kind of underlying effective linear ingredient in the
emission and propagation of the gravitational degrees of freedom
captured in the BBH waveform.

It is important to stress that such effective linearity does not
undermine the fundamental non-linear nature of the basic equations and
the crucial role of non-linearities in BBH
mergers~\cite{sberna2021nonlinear}. Indeed, already the Close-Limit
approximation makes non-linearities apparent by the need of going to
second-order perturbation theory to account for key features of the
merger. But, when dealing with the emitted radiation, BBH merger
waveforms seem to behave more linearly than expected, as seen above.
In this specific setting, the `effective linearity' assumption
proposes 
the effective linear nature of the dominating mechanism underlying the
overall qualitative features of the BBH merger waveform.  The
appropriate background for such linear mechanism to take place should
indeed be controlled by the ultimate non-linear equations, its
identification requiring a good control of the full BBH dynamics.  An
avenue to get some insight into this point is discussed in the
companion article~\cite{JarKri22bv2}. Here, we follow a more agnostic
approach, adopting the `effective linearity' of the mechanism behind
BBH merger waveforms as an assumption.

\subsection{A catastrophe theory approach to BBH mergers: Airy model}

Despite the physical insights and far-sighted nature of the
Close-Limit and the Effective-one-body models, none of these or other
approaches have been able to find a simple functional form for the
merger signal.  This is also reflected in the fact that the
post-Newtonian and Ringdown signals cannot be extended beyond their
regimes of validity (cf. however~\cite{Borhanian:2019kxt}).
  Thus, the signal amplitude in the post-Newtonian
expansion is $(t_c-t)^{-1/4}$ and formally diverges at the coalescence
time $t_c$.  Similarly, the ringdown signals diverge exponentially as
we go backward in time towards the merger.  In essence, the nature of
the gravitational wave signal changes from an oscillatory to a damped
phase and we do not have yet general model which connects these two
asymptotic expansions.

Any analytic description of the merger must necessarily interpolate
between these two radically different signal morphologies.  In this
article we propose such a model for the merger.  We shall argue, based
on ideas from singularity theory developed by V.I.~Arnol'd and
R.~Thom beginning in the 1960s (building  on work by Whitney and
  others), that the merger waveform should indeed
be simple and universal. More concretely, that the Airy function
should be the basis for describing the inspiral-merger transition.
The arguments presented here do not constitute a proof starting from
first principles, but rather a working hypothesis which needs to be
confirmed (or disproved) in further work.  In any case, starting with
this hypothesis, we shall see that at leading order, all
inspiral-merger waveforms can be obtained by a modulated and
parameterized Airy function.

From a more systematic perspective, simplicity is addressed in terms
of an asymptotic reasoning involving two layers: i) a first one in a
`geometric optics' treatment at large frequencies focusing on the
formation of caustics, and ii) a second one incorporating wave
features, namely diffraction over caustics, at an intermediate level
not involving the full (non-linear) wave theory. Universality then
relies on the `structural stability' results for stable
low-dimensional singularities, following from Arnol'd-Thom theorem in
singularity theory. In particular, this theorem classifies the stable
low-dimensional caustics, as well as the (universal) diffraction
patterns over them.  By construction, the proposed framework is
stable under small changes in the dynamical equations, and therefore
will also hold for small modifications to general relativity.

We will begin our discussion with an outline of the Arnol'd-Thom
theory of structurally stable singularities of differentiable
mappings, sometimes referred as Catastrophe theory.  We shall take two
calculations of caustics from optics as prototypical examples for low
dimensional mappings. We shall review how the theory of caustics fits
into this framework, namely the cylindrical lens and the formation of
rainbow.  The cylindrical lens is an example of a cusp-catastrophe,
while the rainbow is a fold-catastrophe.  We shall then argue that a
BBH merger can be viewed as a fold-catastrophe, which will lead us
directly to the Airy function ansatz. Some simple implications of this
ansatz will be considered and finally we shall suggest some directions
for future work. Our goal here is to introduce the framework of
Singularity theory and Caustics to gravitational wave researchers, and
to describe the basic elements of our new model for compact binary
mergers.  Further developments and details necessary for applying
these ideas to gravitational wave data analysis will be discussed
elsewhere.

\section{Caustics in geometric optics: the skeleton for a BBH merger
  model}
\label{s:caustics_skeleton}

The most evident feature of a BBH merger waveform is that shortly
after the merger time $t_{\rm merger}$, its amplitude rapidly
decreases and soon becomes too small to be detectable.  Thus, we might
say that at times before $t_{\rm merger}$, the gravitational wave
detector is ``illuminated'' by gravitational waves, and after
$t_{\rm merger}$ it is not.  Such dramatic changes in illumination are
frequently and commonly observed in optics~\footnote{Caustic phenomena
  are, of course, not exclusive from optics and occur generically in
  the high-frequency (geometric optics) limit of wave theories, such
  as acoustics, seismic fields... For concreteness, we focus on the
  comparison with optics.} and, in particular, occur when passing
through caustics (where the field intensity formally diverges).
Moreover, properties such as the diffraction patterns of the
electromagnetic wave field near caustics have a universal behavior as
we shall see.  Let us then take, as a working hypothesis, that the
behavior of the BBH waveform near $t_{\rm merger}$ is described by an
underlying caustic mechanism.  Before proceeding to investigate the validity and
implications of this hypothesis, it will be useful to review the
standard theory of caustics in geometric optics, and its description
within the framework of Arnol'd-Thom Singularity theory.

\subsection{The rainbow and the lens}
\label{s:caustics}

In the geometric optics approximation to wave propagation at large
frequencies, a propagating wave field corresponds to a family of light
rays, and in our case, "gravitational rays". The determination of the
trajectory from a source to a detector of a given ray is obtained,
according to Fermat's principle, from the extremalization of an
appropriate function, the phase difference $\Phi$. In the case of
monochromatic light this corresponds to the time delay
$t = \Phi/\omega$. Given a set of parameters $\w{X}$ (also referred to
as ``control parameters'') labeling points where a detector could be
placed, the extremalization is performed with respect to a set of
"state variables" $\w{R}$ that encode the physical features of the
source as well as the medium in which the ray propagates.  The space
of state variables will be denoted $S$ and the space of control
variables will be denoted $D$.  The phase difference (`generating')
function $\Phi$ is then a real valued function on $S\times D$.

If a single extremum (`critical point') of $\Phi$ exists for a given
point $\w{X}_o$ in $D$, then a single ray reaches the detector at
$\w{X}_o$ corresponding to a point $\w{R}_o$ in $S$, that
characterizes the ray.  In this simple case, the point $\w{X}_o$ in
$D$ uniquely determines a point $\w{R}_o$ in $S$, a single-valued
function $\w{R}=\w{R}(\w{X})$ can be defined and no caustics appear.
However, if several extremal points of $\Phi$ exist for a given
$\w{X}_o$, i.e. that point $\w{X}_o$ is reached by more than one ray,
then the mapping $\w{X}\mapsto \w{R}$ is multi-valued.  In this
situation, we have the possibility that as $\w{X}_o$ is varied, two or
more extrema of $\Phi$ can coalesce at a given $\w{X}_c$, that defines
a point of the caustic set in the control space $D$: on one side of
the caustic several rays reach $\w{X}_o$, whereas (some of) these rays
disappear at the other side of the caustic.  Within the geometric
optics approximation, the light intensity diverges in the transition
at the caustic.  In reality of course (i.e. at finite frequencies away
from the geometric optics limit), there is no divergence but there is
nevertheless a sharp increase in intensity at the caustic.  Such
dramatic change from "illumination" to "no-illumination" at a caustic
is the mechanism that we intend to explore as providing the `skeleton'
for our model of the BBH merger waveform, namely addressing the
transition from the inspiral phase to the extinction of the signal.

Before proceeding further, we briefly illustrate the discussion above
with two representative examples from optics.

\subsubsection{Example 1: cylindrical lens.}
\label{subsec:lens}

\begin{figure}
\centering
\includegraphics[width=70mm]{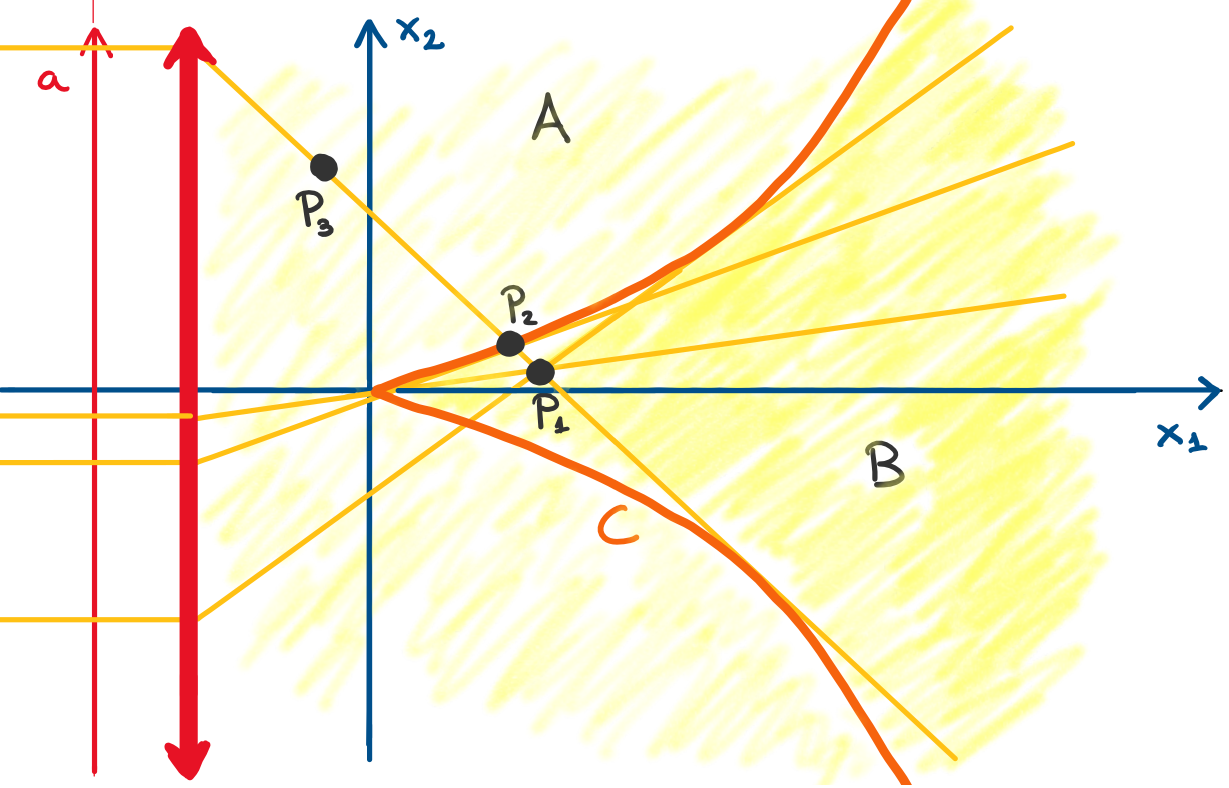}
\caption{Cusp caustic for an axisymmetric lens.  Each point in the
  region $A$ has one ray passing through it while each point in $C$
  has three rays passing through it.  The caustic $C$ separates these
  two regions. }
\label{fig:Axisymmetric_lens}
\end{figure}

We first consider an axisymmetric lens, as in
Fig. \ref{fig:Axisymmetric_lens} (we follow closely
Ref.~\cite{ThoBla17}).  Due to the axisymmetry, the problem reduces to
two-dimensions and we choose to work in the $(x_1,x_2)$ plane with the
incident rays traveling along the $x_1$-axis, and the lens is aligned
along the $x_2$-axis.  A bunch of rays parallel to the $x_1$ axis reaches
the lens, each ray arriving at a distinct height $a$ on it.  This $a$
represents the ``state variable'' $\w{R}$ in this system and the space
$D$ of ``control parameters'' is $\{(x_1,x_2)\}$.  Then, rays are focused
towards the axis of the lens. We assume that the lens is non-ideal in
that the rays do not all meet at the focus, but the rays at a larger
$a$ see effectively a longer focal length.  It can be shown
geometrically that there are three regions $A$, $B$ and $C$ in the
$(x_1,x_2)$ plane: i) only one ray passes through any point in $A$, ii)
three rays pass through each point of $B$ and iii) the curve $C$ with
two components meeting in a cusp, the focus of the lens, separates
$A$ and $B$.  Thus $B$ is more illuminated than $A$ and the
illumination is largest on $C$, which is the caustic (in particular,
the enveloping curve of deflected rays).  Within the geometric optics
approximation, the illumination in fact diverges at the caustic (where
the ray density diverges) and it becomes necessary to invoke the
wave nature of light in order to regularize this divergence.
In more detail, if we move towards the caustic $C$ from its interior
(region B), two of the three rays coincide (``coalesce'') at the
caustic, disappearing when passing to the exterior of the caustic
(region A). Such drastic ray disappearance corresponds to a
singularity at the caustic (in a sense to be specified below), having
as a consequence a `catastrophic' fall of the luminosity when passing
through the caustic.

\begin{figure}
\centering
\includegraphics[width=80mm]{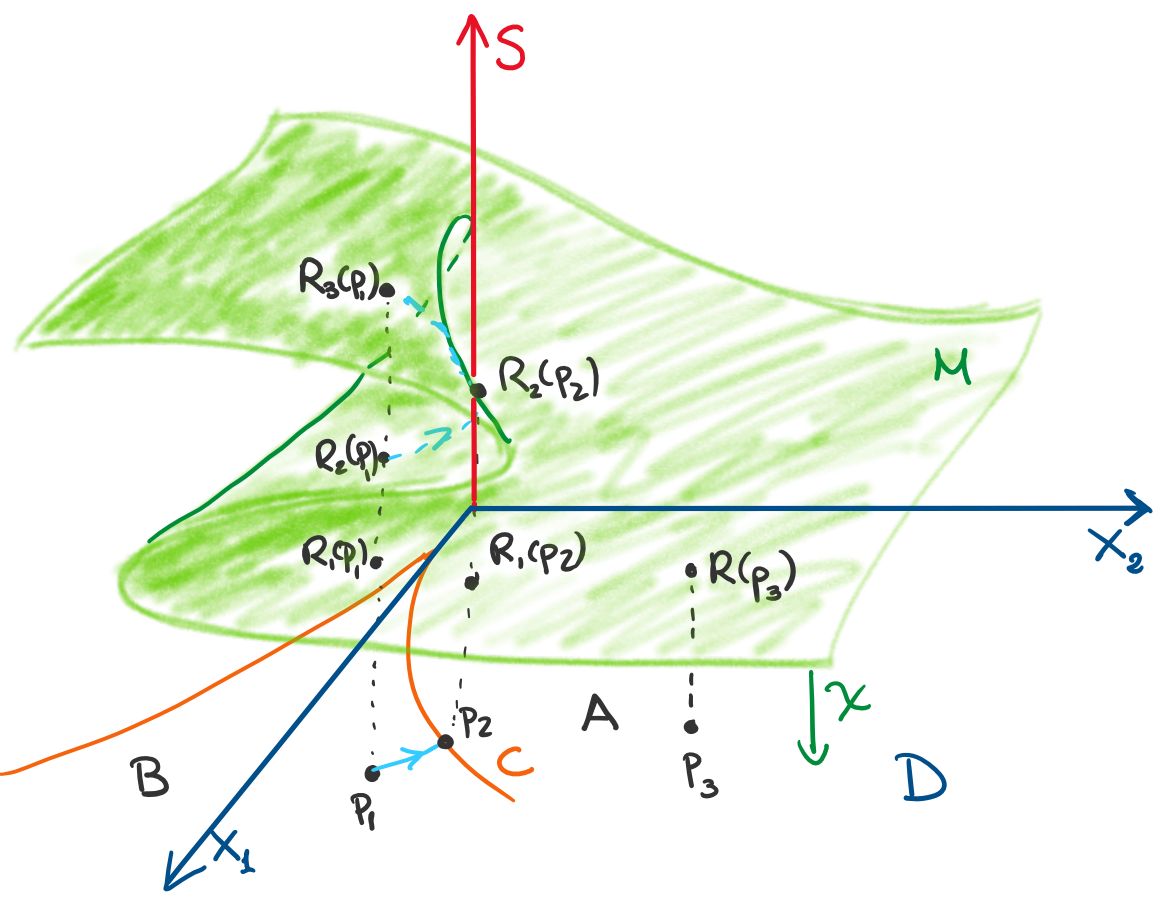}
\caption{The geometry of the cusp catastrophe. The generating
    function $\Phi$ is a quartic polynomial leading to a
    ``stationary manifold'' $M$. The solution space $M$ is a cubic surface
    folding over itself in the space
    $S\times D \equiv \{a\} \times \{(x_1,x_2)\}$. The projection
    $\chi: M\to \{(x_1,x_2)\}$ becomes singular at points where a tangential
    direction to $M$ is vertical (green curve in $M$). The image of
    these points under the projection is the caustic $C$.  The caustic
    separates $D$ in two regions: the interior of the caustic $B$
    where a point $p_1$ has three preimages in $M$ and the exterior of
    the caustic $A$ where a point $p_3$ has only preimage in $M$.
    This explains the higher luminosity in the interior of the
    caustic.  On the caustic, points $p_2$ have two preimages, though
    one of them (resulting from the merger of two stationary points in
    $M$, $R_2(p_2)$ and $R_3(p_2)$ in the figure) is double.}
\label{fig:cusp_geometry}
\end{figure}

Let us revisit the previous discussion in a more systematic manner.
The phase difference $\Phi(a;x_1,x_2)$ is a function of the state and
control variables.  In general this function is multivalued.  For a
given ray, the state variable corresponding to a given point in the
$(x_1,x_2)$ plane is obtained by extremizing the time delay with respect
to the state variable $a$:
\begin{equation}
  \label{e:lens_M}
  \frac{\partial \Phi(a;x_1,x_2)}{\partial a} = 0\,.
\end{equation}
In general, multiple rays will pass through a given point so that this
equation will have multiple solutions.  Solving this equation provides
a set of $a$'s for a given $(x_1,x_2)$ which in turn corresponds to a
hypersurface $M$ in $S\times D=\{(a,x_1,x_2)\}$ space which can ``fold''
over multiple times. If we define the natural projection from $M$ to
$D$, namely $(a,x_1,x_2)\mapsto (x_1,x_2)$, this projections becomes
`singular' precisely at those points where $M$ folds over, their image
in $D$ providing precisely the caustic set $C$
(cf. Fig.~\ref{fig:cusp_geometry}). It is in this sense that the
caustic corresponds to a singularity: it is the image of the singular
points of the projection from $M$ to $D$, namely the ``folding lines''
in the hypersurface $M$ (we illustrate this below).

Alternatively, we can ask ourselves: for which points in $D$ can we
write $M$ locally as a function? Specifically (for a fixed $x_1$), we
can ask ourselves for which $x_2$'s in $D$ we can locally write a
function $a=a(x_2)$.  From the implicit function theorem, this is
determined by the differential of the function
$\frac{\partial \Phi(a;x_2)}{\partial a}$ defining $M$ in
Eq. (\ref{e:lens_M}), more specifically this is controlled by the
second derivative $\partial^2\Phi/\partial a^2$ evaluated at points of
the solution space $M$. ``Branches'' of the manifold $M$ cannot be
locally written as a function in those points of $D$ where this second
derivative vanishes: such points correspond to the caustic $C$.

A straightforward calculation shows that the solution hypersurface $M$
is of the form shown in Fig.~\ref{fig:cusp_geometry} and the
projection onto $D$ is singular at the points when it ``folds'' over.
In more detail, the phase delay function $\Phi$ is a quartic
polynomial in $a$ (cf. e.g.  \cite{ThoBla17})
\begin{equation}
  \label{e:Phi_lens}  
  \Phi_{\mathrm{lens}}(a;x_1,x_2) = \frac{(a-x_2)^2\omega}{2fx_1c}\left(f - x_1 + \frac{a^2x_1}{2s^2}\right)\,.
\end{equation}
where $\omega$ and $c$ are the light frequency and speed,
respectively, $f$ is focal distance and $s$ is a length scale in the
lens. Variables $a$, $x_1$ and $x_2$ are shown in
Fig.\ref{fig:Axisymmetric_lens} with the coordinates $(x_1,x_2)$ centered
at the focal point.

Finding the stationary points of $\Phi$ leads to $M$ as a cubic
surface folding over itself.  Projecting $M$ to the $(x_1,x_2)$ plane, we
see that the projection of the singular points is precisely the
caustic $C$, as commented above.  Finally, it can be shown that the
light intensity $I$ in the geometric optics approximation is proportional
to
\begin{equation}
I\sim \left|\frac{\partial^2\Phi}{\partial a^2}\right|^{-1}_M \,,
\end{equation}
which diverges at the caustic.  This signals the breakdown of this
approximation and it becomes necessary to consider the wave features
of light. As an intermediate stage prior to invoking the full Maxwell
equations, one can consider the Fresnel diffraction
formalism, as we shall shortly discuss.

This example (corresponding to a so-called ``cusp caustic'' due to the
shape of $C$) has the virtue of being intuitive, but it does not
completely capture the transition from ``illumination'' to
``no-illumination'', since rays still reach the outside region. The
following example, although less intuitive, is more faithful to the
actual mechanism that we propose to be at work in BBH mergers, which
we aim to address here.

\subsubsection{Example 2: The rainbow.}\label{subsec:rainbow}

\begin{figure}[t]
\centering
\includegraphics[width=70mm]{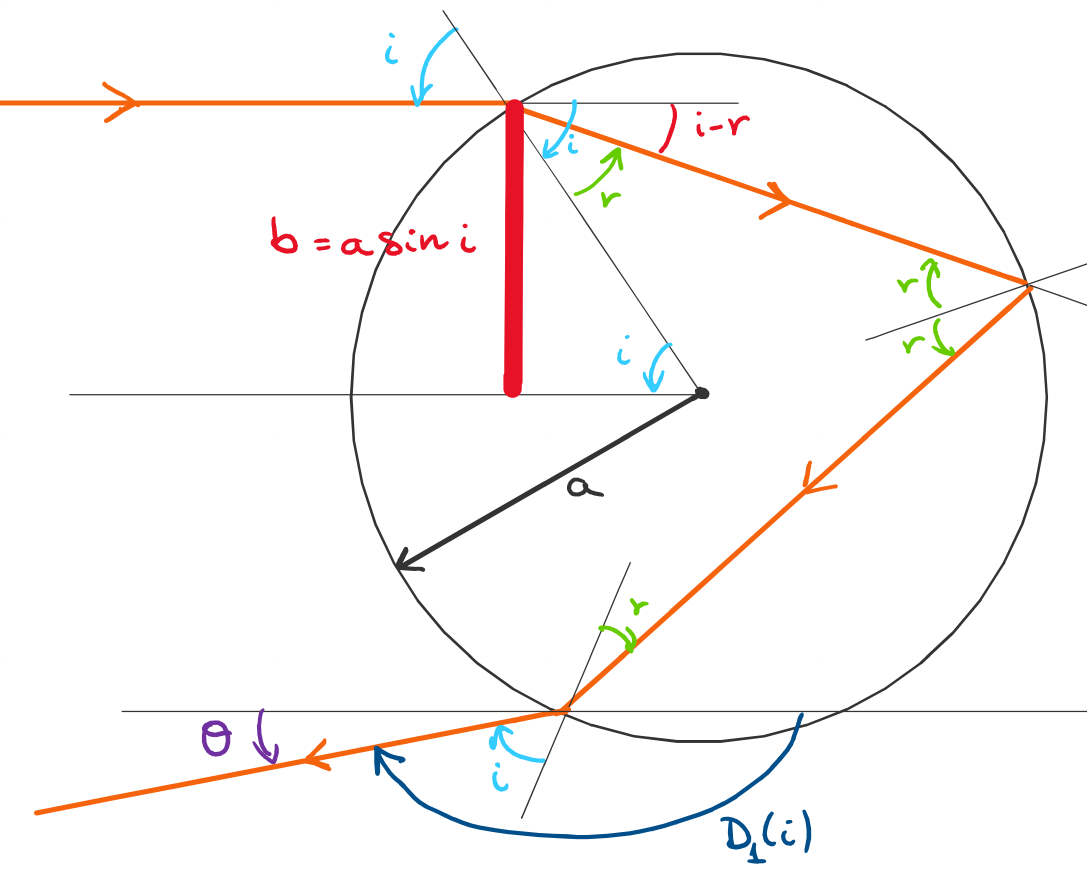}
\includegraphics[width=70mm]{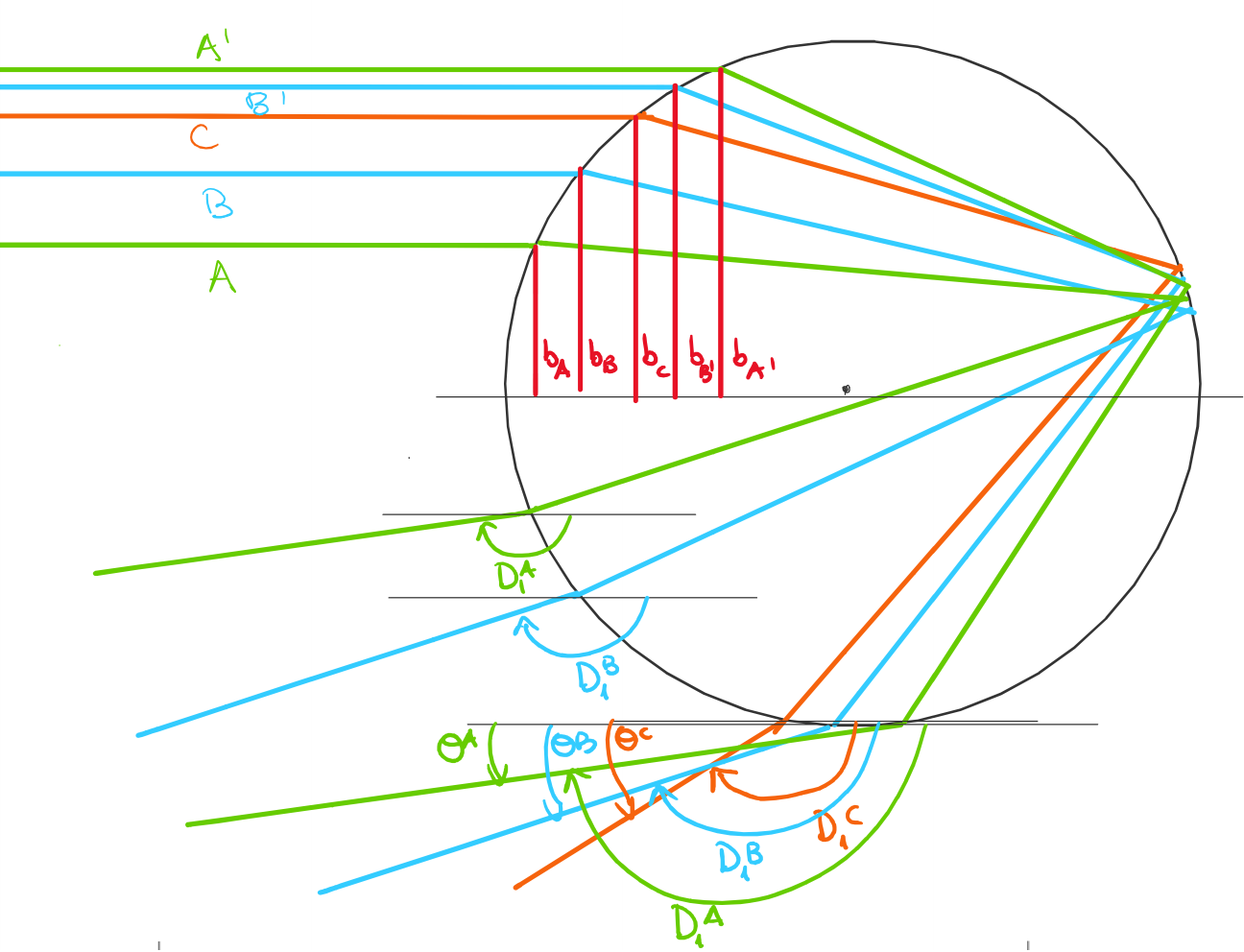}
\caption{Rainbow fold caustic: Parallel incident rays from the left
  are refracted and internally reflected by the water droplet.  There
  is a critical value $\theta_C$ of the deflection angle (depending on the
  wavelength) which is responsible for the rainbow phenomenon. }
\label{fig:Rainbow_fold}
\end{figure}

\begin{figure}[b]
\centering
\includegraphics[width=70mm]{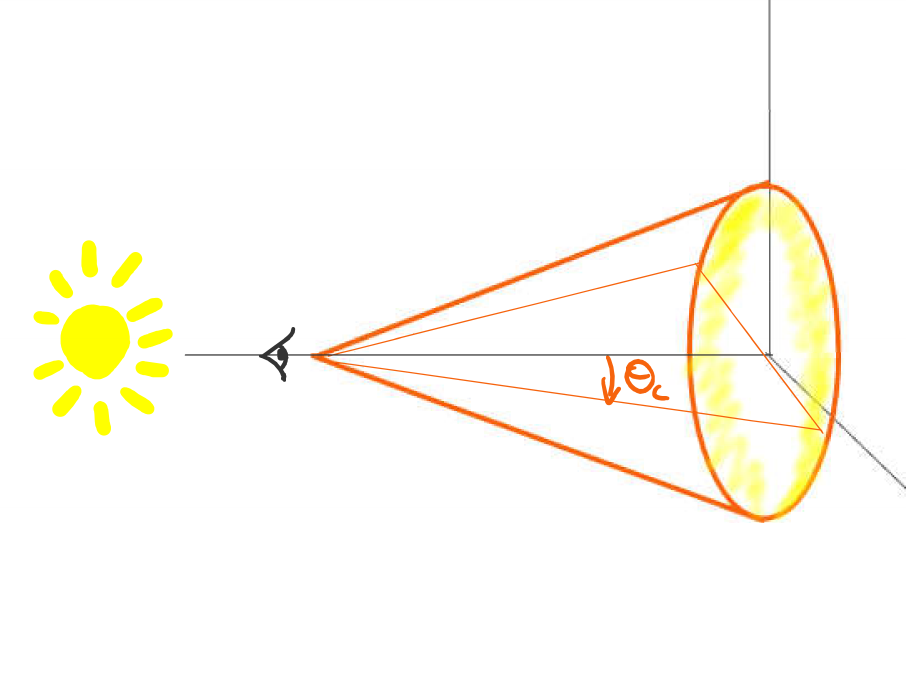}
\includegraphics[width=70mm]{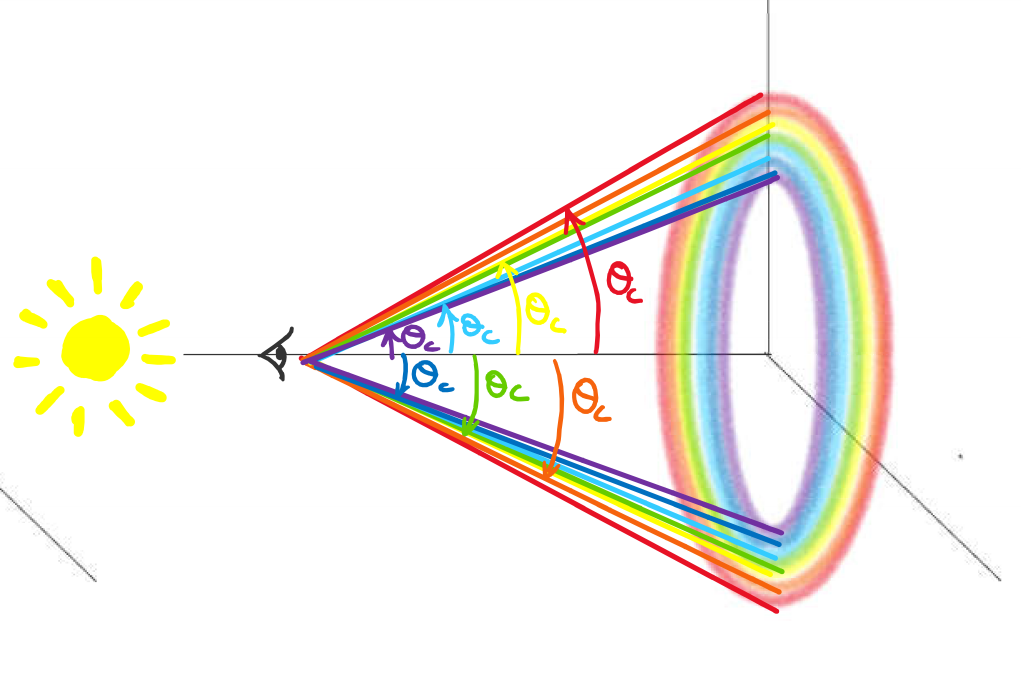}
\caption{Rainbow angular structure. The rainbow corresponds to a caustic
  at angle $\theta_C$, where the density of scattered rays diverges.
  Only the interior of the cone
  with opening angle $\theta_C$ is illuminated (by two rays),
  whereas the exterior of the caustic remains dark.
  In the second panel, the dependence of the
  critical angle on the wavelength is taken into account, resulting in
  the familiar rainbow pattern. }
\label{fig:Rainbow_angular}
\end{figure}

Our second example deals with the physical mechanism of the rainbow
(we closely follow Ref.~\cite{Adam17}). Let us consider a spherical
droplet of radius $a$ and a ray reaching its surface with an incidence
angle $i$.  The light is refracted, internally reflected and finally
exits the droplet after another refraction.  Let the ray enter into
the droplet with a refraction angle $r$. Assuming the refractive index
for air is $n_\mathrm{air}\sim 1$, Snell's law relates the angles as
$\sin i = n \sin r$ with $n$ being the refractive index of water. The
light travels inside the droplet and gets reflected in the internal
side. Finally, the reflected ray reaches again the surface and is
transmitted to the exterior with an angle $i$. Further inner
reflections can be considered and lead to higher-orders
``bows''. Denoting by $D$ the deflected angle with the respect to the
initial incident ray, it holds (cf. first panel of
Fig. \ref{fig:Rainbow_fold})
\begin{equation}
  \label{e:rainbow_deflection_angle}
  D(i) = 2i - 4r(i) + \pi \,.  
\end{equation}
In this problem, one can take as ``state variable'' $\w{R}$ the
incident angle $i$ though, for convenience one can take the ``impact
parameter'' $b=a\sin i$. The natural ``control parameter'' $\w{X}$ is
the angle $D$ or, equivalently, the complementary angle
$\theta = \pi-D$.  Indeed, an observer will receive rays scattered for
all droplets laying in a cone (centered in his or her eye and with
axis pointing away from the Sun) of angle $\theta$. To understand the
rainbow phenomenon, we consider a ray incident perpendicularly at the
droplet, that is $b=0$. This ray is scattered back, i.e. $D=\pi$. As
$b$ increases the deflection angle $D$ diminishes (see angles $D$ of
light rays $A$, $B$ and $C$ in the second panel of
Fig. \ref{fig:Rainbow_fold}). The observer receives rays coming from
all these directions. However, at a critical $b_C$ the deflected angle
reaches a minimum $D^C$, so that when $b$ grows over $b_C$ the angle
$D$ grows. This means that the observer does not receive any ray with
$D<D^C$ (or equivalently, with $\theta>\theta_C$).  Furthermore, the
``density'' of rays accumulated in this turning point $D^C$ formally diverges,
so the associated light intensity at this angle also diverges, i.e. we find a
caustic in the control space at $D^C$.  The rainbow, for a fixed wavelength, is the
arch seen by the observer corresponding to the light received from
droplets along the cone of angle $\theta_C$, for which an enhanced
intensity occurs.  Taking into account light dispersion in water,
namely the fact that blue/violet colors are more scattered (larger
$n$) than red colors, different bows forms for different angles,
with $\theta_C^{\mathrm{blue}}<\theta_C^{\mathrm{red}}$, giving rise
to the familiar rainbow structure (see right panel of
Fig. \ref{fig:Rainbow_angular}).

\begin{figure}
\centering
\includegraphics[width=75mm]{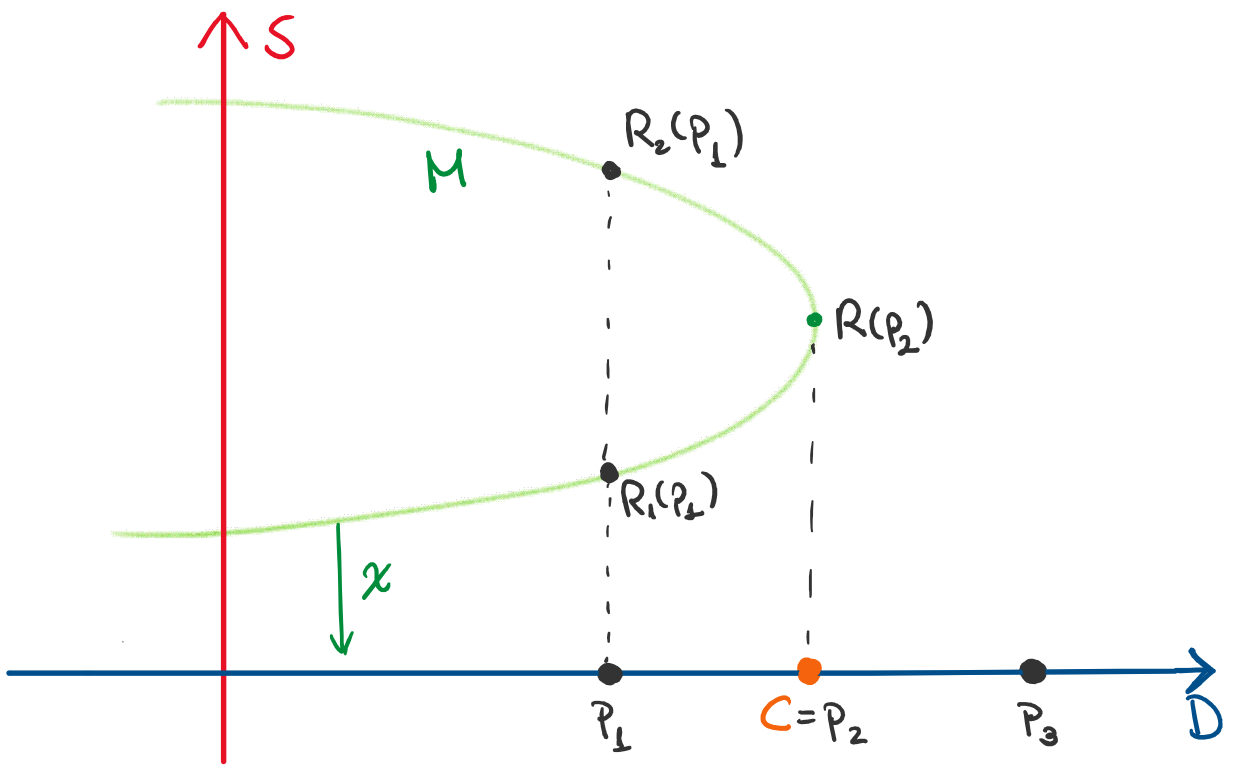}
\caption{Fold catastrophe: The generating function in this case is
  cubic, leading to a quadratic curve in the space (plane)
  $S\times D$. In this case the caustic $C$ is just a point, the image
  under $\chi$ of the vertex of the parabola.  ``Interior'' points
  $p_1$ to the caustic (to the left of $C$ in the figure) have two
  preimages $R_1(p_1)$ and $R_2(p_1)$ in $M$, whereas ``exterior''
  points $p_2$ have no preimage at all. Actually, points $R_1(p_1)$
  and $R_2(p_1)$ in $M$ merge at the caustic, disappearing on the
  other side: in the fold, we pass from illumination to complete
  absence of illumination at the caustic. The rainbow caustic is of
  this type \cite{Adam17}, where the control variable $X_1$ is
  the deflection angle $D$ (cf. Fig. \ref{fig:Rainbow_fold}).}
\label{fig:Fold_geometry}
\end{figure}

From Eq. (\ref{e:rainbow_deflection_angle}) one can derive an
expression for the corresponding phase difference as~\cite{Adam17}
\begin{equation}
  \label{e:Phi_rainbow}
  \Phi_{\mathrm{rainbow}}(\delta; \Delta)_{\mathrm{lens}}  = -\frac{1}{3}\alpha\beta\delta^3 + \beta\delta\Delta \,,
\end{equation}
where $\delta = b-b_c$ and $\Delta= D-D_c$ are, respectively, the
state variable and control parameter centered around the (critical)
caustic values, the parameter $\alpha$ is $\alpha=\frac{1}{2}D''(b_c)$ and $\beta$ an
appropriate constant. The corresponding caustic is referred to as a `fold' and it is illustrated in
Fig.~\ref{fig:Fold_geometry}.

The rainbow is thus a caustic, namely the `fold caustic', in the angle
control parameter $\theta$.  If we consider now the issue of the
illumination at both sides of the caustic, we observe that in the
inner part of the rainbow ($\theta<\theta_C$) there are two scattered
rays reaching the observer, whereas in the outer part
($\theta>\theta_C$) there is no scattered ray at all, so the exterior
is actually dark.  This is indeed observed in actual rainbows in the
sky, where the inner part of the rainbow is more illuminated that the
exterior. This is the same phenomenon we saw above in the (cusp)
caustic of the lens, where couples of rays coalesce at the caustic
(where the intensity diverges) and disappear on the other side. In the
lens (cusp) case no full darkness is found in the exterior due to the
persistence of a third ray.

Our BBH problem, with a passage from an oscillating signal to its total
extinction after the merger (the ringdown is not accounted by this
mechanism) seems mathematically more akin to the rainbow fold caustic
rather than the lens cusp one. We will elaborate later on this point,
once we identify the appropriate ``control parameter'' in our problem,
corresponding formally to the deflecting angle $\theta$.

\subsection{The framework of Singularity theory}

Let us introduce now systematically, though briefly, the relevant
elements in the discussion within the framework of
Singularity/Catastrophe theory.  We shall aim to explain the
abstraction which encompass the examples given above.  Further details
and references can be found in e.g. \cite{Berry76,BerUps80,KraOrl12,Gilmo93,PosSte96,Adam17,ThoBla17}.
\begin{itemize}
\item[i)] {\em Generating function, stationary manifold and
    catastrophe map}.  Let us denote by $S$ a $m$-dimensional space of
  'state variables' $\w{R}=(R_1, \ldots, R_m)$ and by $D$ the
  $n$-dimensional space of 'control' variables
  $\w{X}=(X_1, \ldots, X_n)$.  The a priori number of state variables
  can be large (possibly infinite, as in the case of an underlying
  field theory), but not all of those variables are relevant regarding
  the caustic structure.  The number of independent state variables is
  reduced to the minimal number necessary to encode the (topological)
  structure of the catastrophe. We will focus here on situations with
  a finite, and in fact small, number $m$ of state variables. The
  number of control parameters $n$ could also be infinite
  \cite{Berry76}, but we focus here on a finite number $n\geq m$.  The
  important dimensionality is the dimension of the space of control
  parameters $C$ minus the dimension of the singularity, known as the
  ``co-dimension''. For both the fold and cusp catastrophes the
  singularities are point-like (i.e. the fold and the tip of the cusp);
  thus the fold and cusp catastrophes have co-dimension 1 and 2
  respectively, coinciding with the dimensions of the control space.

  As in the examples presented above, the ray trajectories are
  obtained from the extremalisation of a `generating function',
  $\Phi: S \times D \to \mathbb{R}$ with the structure:
  \begin{equation}
    \label{e:generating_function}    
    \Phi(\w{R}, \w{X}) = \sum_{\alpha=1}^m R_\alpha X_\alpha + f(\w{R};
    X_{m+1}, \ldots,X_n) \,.
  \end{equation}
  Extremizing with respect to the state variables
  \begin{equation}
    \label{e:critical_Phi}
    \nabla_{R_\alpha}\Phi =0\,,\qquad  \alpha\in\{1, \ldots, m\}  \ ,    
  \end{equation}
  permits us to write
  \begin{equation}
    \label{e:X=X(R)}    
    X_\alpha = - \nabla_{R_\alpha}f\,,\qquad \alpha\in\{1, \ldots, m\}\,.
  \end{equation}
  These $m$ relations determine an $n$-dimensional manifold
  $M\subset S\times D$ of stationary points, the latter corresponding
  to actual rays reaching from the source to the detector.  We will
  refer to $M$ as the {\em stationary or catastrophe manifold}
  (cf. e.g. \cite{PosSte96}).  We can define a projection map
  $\chi: M\to D$ (`catastrophe map') as the restriction to $M$ of the
  projection $\pi: S\times D\to D$
  \begin{equation}
    \pi(\w{R},\w{X}) = \w{X} \,,
    \quad \chi=\left.\pi\right|_M \,.    
  \end{equation}
  We can locally parameterize $p\in M$ as
  $p=(R_1,\ldots, R_m, X_{m+1}, \ldots,X_n)$, by using
  Eq.~(\ref{e:X=X(R)}), so 
  \begin{eqnarray}
    \label{e:chi_param}
    && \chi(R_1,\ldots, R_m, X_{m+1}, \ldots,X_n) \\
    &=&\left(X_1(\w{R}),
    \ldots, X_m(\w{R}), X_{m+1}, \ldots,X_n\right) \nonumber
  \end{eqnarray}
\item[ii)] {\em Magnification matrix and intensity}.  If the map
  $\chi$ is one-to-one, for each chosen value of the control parameter
  $\w{X}_o$ it corresponds a unique ray with a given (source) state
  variable value $\w{R}_o$ and, in particular, a single-valued
  $\w{R}=\w{R}(\w{X})$ exists.  We can define the {\em magnification}
  matrix ${\cal M}_{\alpha\beta}$, with
  $\alpha, \beta\in\{1, \ldots,m\}$, of the rays from $S$ to $D$ as
  \cite{Berry76,ThoBla17}
  \begin{eqnarray}
    \label{e:magnification}    
    {\cal M}_{\alpha\beta}(\w{X}) &=& \left(\frac{\partial R_\alpha}{\partial X_\beta}\right)_{( X_{m+1}, \ldots,X_n)}\\ 
    &=&  \left(\frac{\partial X_\alpha}{\partial R_\beta}\right)^{-1}_{ (X_{m+1}, \ldots,X_n)}(\w{R}(\w{X}))  \,.
  \end{eqnarray}
  Using Eqs. (\ref{e:generating_function}), (\ref{e:critical_Phi}) and
  (\ref{e:X=X(R)}) one can write
  \begin{eqnarray}
    \label{e:magnification_hessian}    
    {\cal M}^{-1}_{\alpha\beta}(\w{X}) &=& -\left(\frac{\partial^2 f}{\partial R_\alpha\partial R_\beta}\right)_{(X_{m+1}, \ldots,X_n)}(\w{R}(\w{X})) \\
    &=&  -\left(\frac{\partial^2 \Phi}{\partial R_\alpha\partial R_\beta}\right)_{( X_{1}, \ldots,X_n)}(\w{R}(\w{X}))
  \end{eqnarray}
  The intensity $I$ of the light at the point $\w{X}$ in this
  (geometric optics) ray picture is then given by
  \begin{equation}
    \label{e:intensity_single_valued}
    I(\w{X})= |\mathrm{det} {\cal M}_{\alpha\beta}|\,.
  \end{equation}

\item[iii)]{\em `Branches' of the stationary manifold $M$}.  In generic situations the
  manifold $M$ folds over in $S\times D$ making the projection $\chi$ not
  injective, several `rays' arrive at $\w{X}$ associated with
  different state configurations in the source, i.e. several state
  variables $\w{R}_i\in S$ are in general associated with a given
  $\w{X}\in D$ and the application $\w{R}_i=\w{R}_i(\w{X})$ is
  multi-valued. In this situation, to get the total intensity
  we must correct expression (\ref{e:intensity_single_valued}) 
 by summing over all rays  reaching $\w{X}$, so we must write
  \begin{equation}
    \label{e:intensity_multiple_valued}    
    I(\w{X})= \sum_i |\mathrm{det} {\cal M}_{\alpha\beta}|(\w{R}_i(\w{X})) \,.
  \end{equation}  
  In this setting, we can consider the different branches $S_i$ of
  an {\em action function}  $S_i: D \to \mathbb{R}$ introduced as
  \begin{equation}
    \label{e:action_function}    
    S_i(\w{X}) = \Phi(\w{R}_i(\w{X}), \w{X}) \,.
  \end{equation}
  This defines the different branches of a (generically) multi-valued
  function solving a Hamilton-Jacobi equation \cite{BerUps80} and
  whose constant-value surfaces constitute the geometric wavefronts in
  geometrical optics~\footnote{In geometric optics one usually starts
    actually from the action function, whose gradient provides the
    rays' wave vectors.  This is in general a multivalued function on
    $D$, the different branches corresponding to the folding of $M$ in
    $S\times D$.  However in catastrophe theory this
    `multi-valuedness' is by-passed by introducing the generating
    function $\Phi$ depending on the additional state variables to
    render it a single-valued function, in particular non-singular in
    its arguments.}.

\item[iv)] {\em Caustic}.  The caustic $C$ is precisely the set of
  points in $D$ where the different branches (\ref{e:action_function})
  of a multivalued `action function' join. In other words, the caustic
  is given by the projection to $D$ of the set of points in
  $S\times D$ where $M$ ``folds over itself''. Thus, two stationary
  points $p_i$ and $p_j\in M$ that are separated on one side of the
  caustic, do coalesce $p_i\to p_j$ at the caustic and disappear on
  the other side.

  Technically, the caustic $C$ is the image by $\chi$ of the critical
  points of such projection $\chi$, also referred to as `critical set
  of $M$' (namely the points in $M$ where $d\chi$ is singular,
  i.e. where it has a non-trivial kernel).  From the parametrization in
  (\ref{e:chi_param}) this happens when the Hessian of $\Phi$
  (evaluated on $M$) is not
  invertible. 
  Equivalently, from Eq.~(\ref{e:magnification_hessian}), this happens
  when the inverse of the magnification matrix
  ${\cal M}_{\alpha\beta}$ is singular and its determinant diverges.
  As a consequence of Eq. (\ref{e:intensity_single_valued}) the
  `light' intensity $I$ diverges at the caustic.
\end{itemize}

\subsection{Universal diffraction pattern on caustics}
\label{s:diffraction_caustic}

As we have just seen, the geometric optics approximation to light
propagation breaks down at the caustic, where the intensity
diverges. Moreover, in our BBH setting we are fundamentally
interested, apart from the intensity of the gravitational signal, also
on the radiation wave field at the caustic.  This requires to go
beyond the geometric optics approximation.  Without having to invoke
the full Maxwell theory, a suitable solution to the problem is
provided by Fresnel diffraction integrals~\footnote{ This
  intermediate layer provides another instance of `asymptotic
  reasoning' in the sense of Batterman~\cite{Batte97,Batte01}, where
  full details are sacrificed in order to unveil the underlying
  relevant mechanism.}.  Specifically, diffraction on caustics brings
about two remarkable outcomes: i) it regularizes the intensity
divergence, and ii) diffraction patterns show universal behavior only
depending on the topological nature of the caustic.

The starting point is the Fraunhofer/Fresnel diffraction integral for
the wave field $\psi(\w{X})$ \cite{Berry76,BerUps80,Duist74,GuillStern90,KraOrl12}
\begin{equation}
  \label{e:Fraunhoffer}
  \psi({\w{X}}) = \left(\frac{k}{2\pi}\right)^{\frac{m}{2}}\int d^m\!\w{R} \; e^{ik\Phi(\w{R},\w{X})}a(\w{R},\w{X})\,.  
\end{equation}
Here $k= 2\pi/\lambda$ with $\lambda$ the wavelength,
$\Phi(\w{R},\w{X})$ is an appropriate generating function
(\ref{e:generating_function}) accounting for the phase delay and
$a(\w{R},\w{X})$ is a slowly varying function in $(\w{R},\w{X})$
corresponding to an attenuation factor along the ray path.

Let us consider first diffraction when we are away from a caustic, so 
stationary points $p_i$ of $\Phi(\w{R},\w{X})$ are not degenerate.  In
this situation, we can approximate Eq.~(\ref{e:Fraunhoffer}) using a
{\em stationary phase approximation}.  The phase of the exponential in
(\ref{e:Fraunhoffer}) oscillates rapidly and the main contribution to
the integral comes from the stationary points of $\Phi$.  Expanding up
to second order around a given stationary point $\w{R}_i$ we write
\begin{eqnarray}
  \label{e:Phi_quadratic_expansion}
  \Phi(\w{R},\w{X}) &=&\Phi(\w{R}_i(\w{X}),\w{X})  + \left.\nabla_{R_\alpha} \Phi(\w{R},\w{X})\right|_{\w{R}_i(\w{X})} R_\alpha \nn\\
&+& \left.\nabla_{R_\alpha} \nabla_{R_\beta}
  \Phi(\w{R},\w{X})\right|_{\w{R}_i(\w{X})} R_\alpha R_\beta+
O(\w{R}^3) \ .   
\end{eqnarray}
The linear term vanishes since we are at a stationary point.
Neglecting terms beyond quadratic ones reduces (\ref{e:Fraunhoffer})
to a Gaussian integral contribution around each stationary point
\begin{equation}
  \label{e:Gaussian_integral}
  \psi(\w{X}) \sim \sum_i \frac{e^{i\left(kS_i(\w{X})+\alpha_i
        \pi/4\right)}}{|\mathrm{det}({\cal
      M}^{-1}_{\alpha\beta}(\w{X}))|^{\frac{1}{2}}}a(\w{R}_i(\w{X}),\w{X})  \ ,
\end{equation}
where $\alpha_i$ is the signature (i.e. the number of positive minus the
number of negative eigenvalues, counting multiplicities) of the
${\cal M}^{-1}_{\alpha\beta}$ matrix, and
$S_i(\w{X}) := \Phi(\w{R}_i(\w{X}), \w{X})$ is the action function
in the branch corresponding to $p_i$.  Taking into account that
the field intensity satisfies $I\sim |\psi(\w{X})|^2$, from this
expression we read:
\begin{itemize}
\item[i)] For a single stationary point we recover (if ignoring the
  attenuation factor) the geometric optics value
  (\ref{e:intensity_single_valued}), namely
  $I\sim|\mathrm{det}({\cal M}_{\alpha\beta})|$.
\item[ii)] When multiple non-degenerate stationary points exist, the
  geometric optics expression (\ref{e:intensity_multiple_valued}) is
  corrected with additional interference terms controlled by the
 the action function  evaluated on the stationary points.
\item[iii))] At the caustic the intensity diverges, since
  $\mathrm{det}({\cal M}^{-1}_{\alpha\beta}(\w{X}))=0$. The stationary
  phase approximation is not valid here, and it is necessary to go to
  higher orders in the expansion (\ref{e:Phi_quadratic_expansion})
  and/or modify the diffraction treatment.
\end{itemize}

\subsubsection{Diffraction on caustics: `uniform asymptotic approximation'}

As we have just seen, when stationary points $p_i$ merge at the
caustic the direct stationary phase approximation is no longer valid
and must be modified. A `uniform asymptotic approximation', valid both
for isolated and merged extrema has been systematically developed for
different caustics starting with the work of Chester, Friedman and
Ursell on the fold catastrophe \cite{CheFriUrs57}, later extended to
higher caustics (see references in
\cite{Berry76,BerUps80,KraOrl12}). The subject was put on a rigorous
basis by Duistermaat \cite{Duist74} (see also \cite{Maslo72,Kravt68}).
The method depends critically on the topological
features of the process of merging of stationary points $p_i$ in $M$.
The remarkable fact is that, for finite co-dimension $n$, the {\em
  structurally stable} topological possibilities are classified by the
celebrated Arnol'd-Thom theorem in singularity theory
\cite{wassermann1974stability,broecker1978differentiable}.  Here
``structurally stable''refers to stability under generic small
perturbations.  In the context of caustics in optical theory, that we
shall discuss further shortly, the concept of structural stability
represents effects such as deformations of water droplets due to
gravity, propagation effects in the surrounding medium etc.  These
effects make the physical situation non-ideal but do not change the
essential topological features of the caustic itself.

As an application, this theorem classifies structurally stable
caustics in terms of the generating function $\Phi(\w{R},\w{X})$, that
can be reduced through appropriate re-definitions of state variables
and control parameters to standard universal forms
$\phi(\w{r},\w{x})$.  Table \ref{t:caustic_clasification} presents all
the possibilities for co-dimension $n\leq 3$.  Each of these standard
universal forms is a polynomial in the state and control variables; it
consists of a ``germ'', i.e. a polynomial in the state variable only
which contains the most singular part.  The rest of the polynomial is
linear in the control variables and describes the ``unfolding'' of the
singularity away from the germ.
\begin{table*}
\begin{center}
\begin{tabular}{|c|c|c|}
\hline
Co-dimension & Caustic name & Generating function $\displaystyle \phi(\w{r},\w{x}) $ \\
\hline
$1$ & Fold &   $\displaystyle  r_1 x_1 + r_1^3$\\ 
\hline
$2$ & Cusp &  $\displaystyle r_1 x_1 +\frac{1}{2} x_2 r_1^2 + \frac{1}{4} r_1^4$ \\
\hline 
$3$ & Swallow tail  &  $\displaystyle r_1 x_1 +\frac{1}{2} x_2 r_1^2+\frac{1}{3} x_3 r_1^3 + \frac{1}{5} r_1^5$\\
\hline 
$3$ & Elliptic umbilic &  $\displaystyle -r_1 x_1-r_2 x_2 - x_3 (r_1^2+r_2^2) - 3r_1r_2^2+ r_1^3$\\
\hline 
$3$ & Hyperbolic umbilic &  $\displaystyle -r_1 x_1-r_2 x_2 + x_3 r_1r_2  + r_1^3  + r_2^3$\\
\hline
\end{tabular}
\caption{``Elementary catastrophes'' classifying, in particular,
  structurally stable caustics with co-dimension $n\leq 3$.
  Generating functions $\phi(\w{r},\w{x})$ share the form of
  $\Phi(\w{R},\w{X})$ in (\ref{e:generating_function}), with state
  variables $\w{r}=(r_1, r_2, \ldots, r_m)$ and control parameters
  $\w{x}=(x_1, x_2, \ldots, x_n)$, with $m\leq n$, but lower-case
  letters are chosen to emphasize the fact that these are standard
  universal forms to which equivalent $\Phi(\w{R},\w{X})$ can be
  reduced under appropriate diffeomorphims.}
\label{t:caustic_clasification}
\end{center}
\end{table*}
In this setting, the above-commented `uniform asymptotic
approximation' relies on the comparison of the integral expression
(\ref{e:Fraunhoffer}) for $\psi(\w{X})$ with an integral
\begin{equation}
  \label{e:caustic_diffraction_patterns}
  j(\w{x}) =  \left(\frac{k}{2\pi}\right)^{\frac{m}{2}}\int d^m\!\w{r} \; e^{ik\phi(\w{r},\w{x})} \  ,  
\end{equation}
defined in terms of generating functions $\phi(\w{r},\w{x})$ that
share with $\Phi(\w{R},\w{X})$ the topological structure of the
process of ``coalescence of stationary points'', but such that the choice of state
variables $\w{r}$ and control parameters $\w{x}$ make the functional
form of $\phi(\w{r},\w{x})$ simpler than that of
$\Phi(\w{R},\w{X})$. Under the light of the Arnol'd-Thom theorem, a
natural choice for $\phi(\w{r},\w{x})$ is given by those ones
corresponding to ``elementary catastrophes'' in Table
\ref{t:caustic_clasification}.

As a consequence, the Arnol'd-Thom theorem
not only classifies elementary caustics in geometric optics, but also
the universal diffraction patterns on caustics that ``clothe the
skeleton provided by stable caustics''~\cite{Berry76} when the wave nature of light is
taken into account. In our present setting, if caustics provide the `skeleton'
of the BBH merger waveform model, diffraction can be paraphrased  as providing its `flesh'.

Let us sketch the procedure to express (\ref{e:Fraunhoffer}) in terms
of (\ref{e:caustic_diffraction_patterns}) by closely following
\cite{Berry76} (we bring attention to the recent work~\cite{torres2022wave} for
the application of these  uniform asymptotic approximations in
a gravitational setting~\footnote{We would like to highlight the
  remarkable timing between the notable work~\cite{torres2022wave} and
  the present one, where uniform asymptotic approximations
  for caustic diffraction have been implemented in a gravitational physics setting,
  in both cases largely  building on pioneering Berry's work~\cite{Berry76,BerUps80,Berry17}.}).  We start by considering a generating function $\Phi$
encoding a singularity structure ``equivalent'' to one of the
``elementary catastrophes'', i.e. equivalent to one of the generating
functions $\phi$ in Table \ref{t:caustic_clasification}. More
precisely, $\Phi$ and $\phi$ are equivalent in the sense (cf. e.g. \cite{PosSte96})
that there
exist i) for each fixed $\w{X}$, a diffeomorphism
$\w{R}\mapsto\w{r}(\w{R})$ from the natural state variables in the
problem $\w{R}$ to the standard ones $\w{r}$, ii) a diffeomorphism
between control parameters $\w{X}\mapsto\w{x}(\w{X})$, and iii) a smooth function
$C(\w{X})$ (the ``shear term'') such that 
\begin{equation}
  \label{e:generating_function_equivalence}
  \Phi(\w{R},\w{X}) = C(\w{X}) + \phi(\w{r}(\w{R}),\w{x}(\w{X})) \ .  
\end{equation}
The function $C$ must be, in particular, smooth at the caustic.
To fix $C(\w{X})$ and $\w{x}(\w{X})$, 
we can use that  critical points of $\Phi$
are mapped onto critical points of $\phi$, which follows 
from the fact that 
$C$ is independent of the state variables.
Then, since for co-dimension $n$ there can be $n+1$ critical
points involved in the coalescence \cite{Berry76}, we can impose
\begin{equation}
  \label{e:critical_point_mapping}   
  \Phi(\w{R}_i(\w{X}),\w{X}) = C(\w{X}) +
  \phi(\w{r}_i(\w{x}(\w{X})),\w{x}(\w{X}))\quad i\in\{1, \ldots,
  n+1\} \,.
\end{equation}
For a fixed $\w{X}$ this provides $n+1$ conditions to
determine the $n+1$ numbers $C(\w{X})$ and
$\w{x}(\w{X})= (x_1(\w{X}), \ldots, x_n(\w{X}))$.  Inserting
Eq. (\ref{e:generating_function_equivalence}) into the diffraction
  integral expression
(\ref{e:Fraunhoffer}) for $\psi(\w{X})$, we can write
\begin{equation}
  \label{e:psi_g}
  \psi(\w{X}) = \left(\frac{k}{2\pi}\right)^{\frac{m}{2}}e^{ikC(\w{X})}\int d^m\!\w{r}\; g(\w{r},\w{X}) \; e^{ik\phi(\w{r},\w{x}(\w{X}))} \,,  
\end{equation}
where
\begin{equation}
  \label{e:g_def}
  g(\w{r},\w{X})=\left|\frac{d\w{R}}{d\w{r}}\right|\!(\w{X})\;a(\w{R}(\w{r}),\w{X}) \, ,  
\end{equation}
with $\displaystyle\left|\frac{d\w{R}}{d\w{r}}\right|\!(\w{X})$ the
Jacobian of the map $\w{r}\mapsto\w{R}(\w{r})$, for fixed $\w{X}$.

Then the function $g(\w{r},\w{X})$ is approximated by exploiting the
fact that the main contribution comes from the stationary points
(since $k\gg 1$). Following~\cite{Berry76}, we first write
\begin{eqnarray}
  \label{e:g_rewriting}
  g(\w{r},\w{X}) &=&  g_0(\w{x}(\w{X})) + g_1(\w{x}(\w{X}))\cdot \nabla_{\w{x}}\phi(\w{r},\w{x}) \nonumber \\
  &+& h(\w{r}(\w{x}(\w{X})))\cdot \nabla_{\w{r}}\phi(\w{r},\w{x}) \ ,   
\end{eqnarray}
with $g_1(\w{x}(\w{X}))$ and $h(\w{r}(\w{x}(\w{X})))$ appropriate
$n$-dimensional and $m$-dimensional vectors, respectively.  The
approximation then consists in estimating the third term by its value
at the critical points \footnote{A better approximation would be
  obtained by expanding the third term in (\ref{e:g_rewriting}) around
  the critical points, keeping higher-order terms.  This leads to an
  asymptotic series (see \cite{CheFriUrs57} for the
  fold-catastrophe).}.  Using  the relation  (\ref{e:critical_point_mapping}) this
last term vanishes and we can write
\begin{equation}
  \label{e:g_approx}
  g(\w{r},\w{X}) \sim  g_0(\w{x}(\w{X})) + g_1(\w{x}(\w{X}))\cdot \nabla_{\w{x}}\phi(\w{r},\w{x}) \,.
\end{equation}
Finally, inserting (\ref{e:g_approx}) into (\ref{e:psi_g}) we can
express $\psi(\w{X}) $ in terms of diffraction caustics integrals $j(\w{x})$
defined in Eq. (\ref{e:caustic_diffraction_patterns})
and their derivatives
\begin{eqnarray}
  \label{e:psi_j}
  \psi(\w{X}) &=& e^{ikC(\w{X})}\bigg( g_0(\w{x}(\w{X})) j(\w{x}(\w{X})) \bigg. \nonumber \\
              &+& \left. \frac{g_1(\w{x}(\w{X}))}{ik}\cdot \nabla_{\w{x}}  j(\w{x}(\w{X})) \right) \,.  
\end{eqnarray}
For explicitly known functions $\Phi(\w{R},\w{X})$ and
$a(\w{R},\w{X})$ a further step can be taken~\cite{Berry76} to determine, for each
$\w{X}$, the $n+1$ numbers $g_0(\w{x}(\w{X}))$ and
$g_1(\w{x}(\w{X}))$~\footnote{Starting from Eq. (\ref{e:generating_function_equivalence}), it
  follows the relation
  \bea
 \label{e:dPhi_dphi}
\nabla_{R_\alpha} \Phi \; dR_\alpha = \nabla_{r_\alpha} \phi \; dr_\alpha \  .
\eea
Then we can estimate $g(\w{r},\w{X})$ in Eq. (\ref{e:g_def}) with the
approximation (\ref{e:g_approx}) and, by expanding (\ref{e:dPhi_dphi})
to second order, that provides
\begin{equation}
  \left|\frac{d\w{R}}{d\w{r}}\right|\!(\w{X}) =
  \left(\frac{\mathrm{Hess}(\phi)(\w{r}(\w{x}(\w{X})))}{\mathrm{Hess}(\Phi)(\w{R}(\w{X}))}\right)^{\frac{1}{2}}\,,
\end{equation}
and the following $n+1$ equations at the critical points
\begin{eqnarray}
  \label{e:fixing_g0_g1}
  &&\left(\frac{\mathrm{Hess}(\phi)(\w{r}_i(\w{x}(\w{X})))}{\mathrm{Hess}(\Phi)(\w{R}_i(\w{X}))}\right)^{\frac{1}{2}}
  a(\w{R}_i(\w{X}), \w{X}) \\
  &=& g_0(\w{x}(\w{X})) + g_1(\w{x}(\w{X}))\cdot \nabla_{\w{x}}\phi(\w{r}_i(\w{x}(\w{X})),\w{x}(\w{X}))  \ ,
\end{eqnarray}
fix $g_0(\w{x}(\w{X}))$ and the $n$-vector $g_1(\w{x}(\w{X}))$, for
each $\w{X}$.  }. Although of genuine interest, we will not need
this further step at the present stage of the discussion of the BBH
merger waveform model.

We  would like to conclude this section with some remarks:
\begin{itemize}
\item[i)] {\em Universal diffraction pattern}. Expression (\ref{e:psi_j}),
   captures the functional form
  of the wave field in terms of universal diffraction patterns on elementary caustics
  (even in the case when functions $\Phi(\w{R},\w{X})$ and $a(\w{R},\w{X})$  are not known).
  This expression, consequence of the `uniform asymptotic approximation',
  is the main result we want to focus on here, as the
key ingredient for a template proposal for BBH merger waveforms.

\item[ii)] {\em Transitional approximation}. The full uniform
  asymptotic approximation is valid on and near the caustic, but also
  far from it. If we focus on the region close to the highest-order
  singularity in the caustic, the second term in (\ref{e:psi_j}),
  involving the derivative of the elementary diffraction, can be
  neglected~\cite{Berry76} (also $\w{x}$ could be taken as
  $x(\w{X})=\w{X}$).  This defines the so-called {\em transitional
    approximation} \cite{ForWhe59}.
  
  On the other hand, if we are interested in connecting with the behavior
  far from the caustic, a stationary phase expansion on the critical
  points of $\phi$ leads to a recovery of the expression
  (\ref{e:Gaussian_integral}) (cf. e.g. \cite{Berry76,BerUps80,KraOrl12,Menes_et_al22}).

\item[iii)] {\em Diffraction catastrophe universal scaling laws}.
  As $k\to \infty$,  the diffraction pattern must recover the geometric optics limit,
  in particular the divergence at the caustic and the vanishing of the spacing in the
  diffraction fringe patterns. This is indeed the case and, moreover,
  such behavior is universal and controlled by scaling laws only depending
  on the caustic type~\cite{Berry77,BerUps80,KraOrl12,Batte01}. Specifically, introducing the {\em elementary diffraction catastrophe}
  \begin{equation}
    \label{e:capital_J}
    J(\w{x})=\left(\frac{1}{2\pi}\right)^{\frac{m}{2}}\int d^m\!\w{r}
    \; e^{i\phi(\w{r},\w{x})} \,,    
  \end{equation}
  not depending on $k$, the asymptotic solution for large $k$
  satisfies the scaling (self-similar) law~\footnote{Such self-similar behavior
  captured by Eq. (\ref{e:scaling_laws}) close
  to the caustic could play in the interplay between self-similarity and
  phase transitions discussed in~\cite{JarKri22bv2} for BBH merger waveforms
  from a PDE perspective.}
  \begin{equation}
    \label{e:scaling_laws}
    j(x)=k^\beta J(k^{\sigma_i}x_i) \ ,
  \end{equation}
  with scaling exponents $\beta$ and $\sigma_i$, with $i\in\{1,\ldots,n\}$ ($n$ being the
  co-dimension of the caustic). The exponent $\beta$, referred to
  as {\em singularity index} and introduced by Arnol'd, controls the divergence
  of the wave amplitude close to the highest order singularity of the caustic
  \begin{equation}
    |\psi(\w{X})| \sim k^\beta \,.
  \end{equation}
  On the other hand, the exponents $\sigma_i$ (introduced by
  Berry~\cite{Berry77}) control the fringe spaces along the control
  variables $x_i$ as $k$ diverges. In particular, the so-called
  `fringe index' $\gamma$ is defined as $\gamma=\sum_{i}^n \sigma_i$
  (cf.~\cite{Berry77}).

\item[iv)] {\em Relevant and irrelevant state variables.} Given a
  control space of dimension $n$, the number of state variables $m$ is
  constrained by $m\leq n$ (cf. e.g. table
  \ref{t:caustic_clasification}).  Additional state variables do not
  change the diffraction catastrophe analysis, since they enter
  quadratically in the generating function $\Phi$ and therefore do not
  contribute at the caustic. In other words, the physical system could
  be described by a large (infinite, in a field theory) number of state
  variables and still the relevant state variables remain finite and
  constrained by the caustic codimension.
\end{itemize}

\section{Fold caustics and binary black hole mergers}
\label{s:fold_caustic_model}

\subsection{Assumptions in the BBH caustic model}
\label{s:BBH_caustics_assumptions}

We are now in a position to formulate the elements of a model for the
BBH merger waveform, based on the diffraction of gravitational waves
on a caustic. The model is built on the following assumptions:
\begin{itemize}
\item[1.] {\em Main Ansatz.} BBH gravitational wave propagation from
  the source to the detector can be modeled in a geometric-optics
  approximation in terms of the extremalization of a phase-difference
  function $\Phi(\w{R},\w{X})$, with control parameters given by the
  spacetime coordinates of the detector. The transition from the
  growing intensity of the signal in the inspiral phase to the
  disappearance of the signal after the merger can be described in
  terms of a caustic (catastrophe).

\item[2.] {\em Co-dimension $n=1$: fold caustic}. Along the world-line
  of the detector, the only relevant control parameter is the time, so
  that $\w{X}=t$. According to Ansatz 1 and the Arnol'd-Thom theorem,
  the catastrophe should be a fold-caustic and, as a consequence: i) there is
  a single
  (relevant) state
  variable $R$ for the whole source system, and ii) the phase
  difference $\Phi(R,t)$ is topologically equivalent to
  $\displaystyle \phi(r,\tau)=\frac{1}{3}r^3 + \tau r$ with $r$
  and $\tau$ obtained by reparametrizations of $(R,t)$.

\item[3.] The natural range of the parameter $r$ is assumed to be to
  full real line, i.e. $r\in (-\infty,\infty)$.  No other assumptions
  are made on $R$ or $r$ at this stage, which remain as abstract
  ``effective source variables''~\footnote{In a further step, it will
    be of interest to explore the relation of $R$ and $r$ with
    physical system parameters, accounting for the BBH caustic
    description from first principles in terms of the evolution of the
    two orbiting black holes in an appropriate point-particle modeling
    of the BBH dynamics~\cite{Menes_et_al22}.}.  Spacetime point labels (namely $t$ here)
  are ``control parameters'', and not ``state variables'', in a spirit
  akin with the general relativistic perspective.
\end{itemize}

\subsection{Diffraction on a fold caustic: the Airy uniform approximation}
\label{s:Airy_uniform}

Under the previous assumptions, we model each of the two polarizations
of the gravitational waveform as the Fraunhofer/Fresnel diffraction
regularization of the fold-caustic geometric optics description of the
merger (cf. the treatment in \cite{KraOrl12} for the
  electromagnetic case). Specifically, we replace $\Phi$ with its
topologically equivalent $\phi$, and begin with the integral
\begin{equation}
  \label{e:j_Airy}
  j(\tau) =  \left(\frac{\omega}{2\pi}\right)^{\frac{1}{2}}\int_{-\infty}^\infty dr \;
  e^{i\omega(\frac{1}{3}r^3 + \tau r)} \,.
\end{equation}
Notice that, for notational reasons, we have substituted the `wave
number' $k$ parameter in Eq. (\ref{e:caustic_diffraction_patterns}) by
the `frequency' parameter $\omega$, to stress that the caustic happens
in time, and not in space.  The diffraction integral $j(\tau)$ in
Eq. (\ref{e:j_Airy}) is directly related to the Airy function, which
is defined as
\begin{equation}
  \label{eq:airydefn}
  \mathrm{Ai}(\tau) = \frac{1}{2\pi}\int_{-\infty}^\infty e^{i(\tau r+r^3/3)}dr\,,
\end{equation}
(notice the relation $\mathrm{Ai}(\tau)=(1/\sqrt{2\pi}) J(\tau)$, with
$J(\tau)$ the universal caustic diffraction pattern in Eq. (\ref{e:capital_J}),
for the particular fold case).
On the one hand, making the change $\omega r^3=\tilde{r}^3$ we can
rewrite the $\omega$-dependent $j(\tau)$ in terms of
$\mathrm{Ai}(\tau)$ as
\begin{eqnarray}
  \label{e:Arnoldindex_fold}
  j(\tau) &=& \omega^\frac{1}{6}\left(\frac{1}{2\pi}\right)^{\frac{1}{2}}\int_{-\infty}^\infty d\tilde{r} \; e^{i( \frac{1}{3}\tilde{r}^3 + \omega^{\frac{2}{3}}\tau \tilde{r})} \nn\\
  &=& \omega^\frac{1}{6}(2\pi)^{\frac{1}{2}}\mathrm{Ai}( \omega^{\frac{2}{3}}\tau) \, .
\end{eqnarray}
On the other hand, considering the BBH phase difference $\Phi(R,t)$ in
Ansatz 2, with appropriate diffeomorphisms $r(R)$ and $\tau(t)$
rendering it topologically equivalent to the canonical fold-generating
function $\phi(r,\tau)$ in Table \ref{t:caustic_clasification}, we can
apply the analysis in section \ref{s:diffraction_caustic} to write each
  gravitational wave polarization $\psi(t)$ in terms of $\mathrm{Ai}$
  and $\mathrm{Ai}'$. Specifically, applying Eq.~(\ref{e:psi_j}) to
  the fold-caustic case, we get
\begin{eqnarray}
  \label{e:Airy_psi_omega}
  \psi(t) &=& e^{i\omega C(t)}\left(\omega^{\frac{1}{6}}\tilde{g}_0(t)
  \mathrm{Ai}(\omega^{\frac{2}{3}}\tau(t))\right. \nonumber\\
          &-& \left. i \omega^{-\frac{1}{6}}\tilde{g}_1(t) \mathrm{Ai}^\prime(\omega^{\frac{2}{3}}\tau(t)) \right) \,.  
\end{eqnarray}
 Two remarks are in order:
\begin{itemize}

\item[i)] {\em Fold-diffraction critical exponents}. Comparing Eq.~(\ref{e:Arnoldindex_fold})
  with the self-similar scalings in Eq. (\ref{e:scaling_laws}) we can identify the
  Arnol'd parameter singularity index $\beta$ and the  fringe index $\gamma=\sigma_\tau$
  for the fold-diffraction, respectively $\beta = \frac{1}{6}$ and $\gamma=\frac{2}{3}$.
  In particular, the singularity index $\beta$ predicts that, in the vicinity
  of the merger,
  the amplitude $|\psi|$ that provides the envelope of the gravitational wave signal
  presents the following power-law dependence in the frequency $\omega$
  \bea
  \label{e:Fourier_transform_close_merger}
   |\psi(\omega)|\sim \omega^{\frac{1}{6}} \ .
  \eea

\item[iii)] {\em Airy pattern beyond `catastrophe optics'}. A tenet in
  asymptotic reasoning~\cite{Batte97,Batte01} is that asymptotic
  limits (in the present case, geometric optics limit
  $\omega\to\infty$) permit to identify the relevant pattern
  underlying the studied physical mechanism, the pattern itself
  remaining often valid away of that limit. This does not mean that
  the identified pattern accounts for the detailed quantitative
  aspects, but rather that it captures the relevant qualitative
  features. In this spirit, we promote expression
  (\ref{e:Airy_psi_omega}) to the following $\omega$-independent
  pattern
  \begin{eqnarray}
    \label{e:Airy_gen}
    \psi(t) = e^{iC(t)}\left(g_0(t)
    \mathrm{Ai}(\tau(t))
    - i g_1(t) \mathrm{Ai}^\prime(\tau(t)) \right) \,.  
  \end{eqnarray}
  This $\omega$-independent pattern will be indeed recovered in a
  completely independent approach in \cite{JarKri22bv2}.  For our
  purposes here, this will be the starting point to be further
  development below which will lead to our model for the gravitational
  wave signal.
  
\end{itemize}

\begin{figure}
\includegraphics[width=\linewidth,trim=0 6cm 0 6cm,clip]{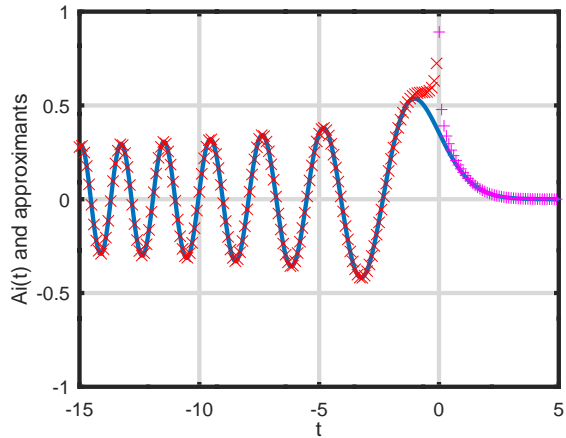}
\caption{The behavior of the Airy function on the real axis.  Also
  shown are the two approximations from Eq.~(\ref{eq:airyapprox1}) for
  negative times (red) and from Eq.~(\ref{eq:airyapprox2}) for
  positive times (magenta).}
\label{fig:airyplot}
\end{figure}
\subsubsection{Some properties of the Airy function}
\label{s:Airy_properties}

We now enumerate some properties of the Airy function that we shall
use below.  We start with the Airy function defined in
Eq.~(\ref{eq:airydefn}) for real arguments and satisfying the equation
$\mathrm{Ai}^{\prime\prime}(t) = t\mathrm{Ai}(t)$.  The Airy function can be extended
as a holomorphic function $\mathrm{Ai}(z)$ over the entire complex
plane satisfying the Equation
\begin{equation}
  \frac{d^2 \mathrm{Ai}(z)}{dz^2} - z\,\mathrm{Ai}(z) = 0\,.
\end{equation}
From a steepest descent evaluation of the Airy function, we have the
following useful asymptotic expansions \cite{erdelyi1956asymptotic}.
For (real) $t$ large and negative, it holds
\begin{equation}
  \label{eq:airyapprox1}
  \mathrm{Ai}(t) \sim \frac{1}{\sqrt{\pi}}\frac{1}{(-t)^{1/4}}\cos\left(
    \frac{2}{3}(-t)^{3/2} - \frac{\pi}{4} \right)  \,.
\end{equation}
We notice the factor $(-t)^{-1/4}$ controlling the asymptotic growth
of the Airy function in this `early' asymptotic limit. In particular,
although the waveform phase requires a reparametrization which is not
determined so far, the familiar amplitude growth as $(-t)^{1/4}$ in
the BBH inspiral phase is exactly captured by Airy asymptotics. As a
matter of fact, already the interference terms in
Eq.~(\ref{e:Gaussian_integral}) account for this behavior through the
determinant of the matrix ${\cal M}_{\alpha\beta}$.  This is in a
tantalizing agreement with the BBH inspiral behavior arising as a
result of combining energy balance and Kepler's laws and, more
generally, post-Newtonian calculations.
  
For positive and sufficiently large $t$, we have
\begin{equation}
    \label{eq:airyapprox2}
  \mathrm{Ai}(t) \sim \frac{1}{2\sqrt{\pi}}\frac{1}{t^{1/4}}e^{-\frac{2}{3}t^{3/2}} \,.
\end{equation}
Thus, the decay is super-exponential for positive $t$.  A plot of the
Airy function on the real axis and these two asymptotic approximations
is shown in Fig.~\ref{fig:airyplot}.  These asymptotic expansions
yield excellent approximations to the Airy functions, and diverge from
the true values only for a small interval around $t=0$.  In fact,
Eq.~(\ref{eq:airyapprox1}) is seen to provide an excellent
approximation even up to the peak of the Airy function, though it
fails to capture the subsequent decay.  The first peak of the Airy
function, i.e. the first zero of $\mathrm{Ai}^\prime(t)$, occurs at
$t_1 \approx -1.019$ and the peak value of the Airy function is
$\mathrm{Ai}(t_1) \approx 0.5357$.  The value of the Airy function at
$t=0$ is $\mathrm{Ai}(0) \approx 0.3550$ (we note that an
inflection point of the Airy function occurs at $t=0$, a feature
of potential interest for later signal modeling).

It will be important in the Airy model below to be able to apply a
phase shift to the Airy function.  To this end, we define the
functions $\mathrm{Ai}_+(z)$ and $\mathrm{Ai}_-(z)$ as follows
\begin{equation}
  \mathrm{Ai}_+(z) = \mathrm{Ai}\left(e^{-2\pi i/3}z\right)\,, \quad
  \mathrm{Ai}_-(z) = \mathrm{Ai}\left(e^{2\pi i/3}z\right) \,.
\end{equation}
It can then be shown that (see e.g. \cite{erdelyi1956asymptotic,TaylorAiry})
\begin{equation}
  \mathrm{Ai}(z) = e^{\pi i/3}\mathrm{Ai}_+(z)   + e^{-\pi i/3}\mathrm{Ai}_-(z)  \,.
\end{equation}
It is also easy to see that
$\overline{\mathrm{Ai}_+({\overline{z}})} = \mathrm{Ai}_-(z)$.  Thus,
along the real axis, where we write $z=t$, we will have
$\overline{\mathrm{Ai}_+(t)} = \mathrm{Ai}_-(t)$ and
\begin{eqnarray}
  \mathrm{Ai}(t) &=& e^{\pi i/3}\mathrm{Ai}_+(t) + e^{-\pi i/3}\mathrm{Ai}_-(t)\nonumber \\
                 &=& 2\Re\left[e^{\pi i/3}\mathrm{Ai}_+(t) \right]\,.
                     \label{eq:AiryPhase}
\end{eqnarray}
We can obtain steepest descent approximations to $A_\pm(z)$. For
example, in the region in the complex plane with
$|\mathrm{Arg}(z) - 2\pi/3| < (1-\epsilon)\pi$ (for some
$\epsilon > 0$), we have
\begin{equation}
 \mathrm{Ai}_+(z) \sim \frac{e^{\pi i/6}}{2\sqrt{\pi}z^{1/4}}
 e^{\frac{-2i}{3}(-z)^{3/2}}  \,.
\end{equation}
Since this is of the form of a complex exponential, the form of the
Airy function given in Eq.~(\ref{eq:AiryPhase}) allows us to define a
phase shift of $\pi/2$ in the Airy function (just as the sine and
cosine functions have a phase offset of $\pi/2$)
\begin{eqnarray}
  \mathrm{Ai}_{-\pi/2}(t) &:=& e^{\pi i/3}e^{-i\pi/2}\mathrm{Ai}_+(t) + e^{-\pi i/3}e^{i\pi/2}\mathrm{Ai}_-(t)  \nonumber\\
  \label{eq:airyphaseshift}
                         &=& 2\Im\left[e^{\pi i/3}\mathrm{Ai}_+(t) \right]\,.
\end{eqnarray}
The following asymptotic expansions for large negative $t$ then
follows
\begin{eqnarray}
  \mathrm{Ai}_{-\pi/2}(t) \sim \frac{1}{\sqrt{\pi}}\frac{1}{(-t)^{1/4}}\sin\left(
    \frac{2}{3}(-t)^{3/2} - \frac{\pi}{4} \right)  \,.
\end{eqnarray}
Comparing with the asymptotic expansion for $\mathrm{Ai}(t)$ given in
Eq.~(\ref{eq:airyapprox1}), we see explicitly the phase shift of
$\pi/2$.  It is straightforward to introduce a general phase shift
$\varphi_0$
\begin{equation}
  \mathrm{Ai}_{\varphi_0}(t) := e^{\pi i/3}e^{i\varphi_0}\mathrm{Ai}_+(t) + e^{-\pi i/3}e^{-i\varphi_0}\mathrm{Ai}_-(t)  \,.
\end{equation}
The phase shifted functions $\mathrm{Ai}_{\varphi_0}(t)$ and
$\mathrm{Ai}_{\varphi_0-\pi/2}(t)$ shall be used below to model the
two polarizations of the gravitational wave signal.

\section{The Airy model for compact binary mergers}

\subsection{Transitional approximation to
  Airy-diffraction binary merger waveforms}
In the previous sections we have argued that the gravitational wave
signal for a generic compact binary merger (excluding the post-merger
phase) should be described in terms of the (amplitude) modulation and
reparameterization of the Airy function and its derivative, as
captured in Eq. (\ref{e:Airy_gen}).  Conventionally, the two
polarizations of the gravitational wave signal from a coalescing
binary are written in terms of reparameterized sine and cosine
functions
\begin{eqnarray}
  \label{eq:strainplus}
  h_+(t) &=& A_+ \eta(t)\cos\varphi(t)\,,\\
  \label{eq:straincross}
  h_\times(t) &=& A_\times\eta(t)\sin\varphi(t)\,.
\end{eqnarray}
Here $\eta(t)$ is a slowly varying function which determines the
variation of the amplitude, while the phase $\varphi(t)$ is a rapidly
varying function of $t$.  For the dominant quadrupole mode
($\ell=m=2$), the amplitudes $A_{+,\times}$ are
\begin{equation}
  A_+ = \frac{1+\cos^2\iota}{2}\,,\quad A_\times = \cos\iota\,,
\end{equation}
where $\iota$ is the inclination of the angular momentum of the binary
with the line-of-sight to the detector.  The parameters of the binary
system appear in $\eta(t)$ and $\varphi(t)$.  Let $m_1$ and $m_2$ be
the masses of the two black holes, $M= m_1+m_2$ the total mass,
$\mu=m_1m_2/M$ the reduced mass, and $\mathcal{M} = \mu^{3/5}M^{2/5}$
the chirp mass.  Let $D$ be the distance to the source.
Then, within the post-Newtonian approximation we have
\begin{equation}
  \eta(t) = \frac{G\mathcal{M}}{c^2 D}\left(\frac{t_c-t}{5G\mathcal{M}/c^3} \right)^{-1/4}\,.
\end{equation}
At the coalescence time $t_c$, the separation between $m_1$ and $m_2$
vanishes, and the amplitude diverges. The phase $\varphi(t)$ depends
on the masses and also on the spins of the two black holes.  At
leading order
\begin{equation}
  \label{eq:pnphase}
  \varphi(t) = \varphi_0 - 2\left(\frac{t_c-t}{5G\mathcal{M}/c^3} \right)^{5/8}\,.
\end{equation}
The coalescence phase $\phi_0$ is the phase when $t=t_c$.  Higher
order corrections including also other parameters such as spin and
eccentricity are well known in the literature.

As we approach the coalescence time, $\eta(t)$ varies increasingly
rapidly and eventually diverges.  Thus, the distinction between the
amplitude and phase as being respectively slowly and rapidly varying
is no longer valid.  As discussed earlier, we postulate the underlying
mechanism of this divergence to be akin to the divergence of the light
intensity at a caustic in the geometric optics approximation.

Summarizing the discussion from sections \ref{s:caustics_skeleton} and
\ref{s:fold_caustic_model}, to regularize this divergence we thus
need to invoke the relevant Fresnel integral with an appropriate
choice of phase function $\Phi$ `topologically equivalent' to one of
the canonical phase functions $\phi$ taken from
Table~\ref{t:caustic_clasification}.  In our case, the relevant
control parameter is just the time $t$ at which the detector collects
data. Since we just have then a single control parameter it follows
from our hypothesis and the Arnol'd-Thom theorem that the function
$\Phi$ must be equivalent to the $\phi$ generating function for a fold
catastrophe, which eventually leads an expression in terms of the Airy
function and its derivative.  We start then with the result of
Eq.~(\ref{e:Airy_gen}), namely a \emph{uniform} approximation to the
diffraction integral.

If we focus on the merger waveform, it is appropriate to consider
the transitional approximation discussed in point ii) of
section \ref{s:diffraction_caustic}, namely we 
drop the second term in Eq.~(\ref{e:Airy_gen}),
involving the derivative of the Airy function \cite{Berry76,ForWhe59}.
Remarkably, in this particular case involving a one-dimensional control space,
it can be justified from the matching of the uniform
approximation with the early (interference of `ray fields') expression
(\ref{e:Gaussian_integral})  valid far away from the caustic,
that the second term in derivatives of the Airy function vanishes
exactly~\cite{KraOrl12,Menes_et_al22}. This permits to extend
the transitional approximation for all the signal and work
with a modulated and reparametrized Airy function.

\subsection{The Airy model for binary mergers}
\label{s:the_model}

In practical terms, by dropping the $\mathrm{Ai}'$ term in
Eq. (\ref{e:Airy_gen}) through the transitional approximation, the
role of the Airy function becomes that of replacing the sine and
cosine in Eqs.~(\ref{eq:strainplus}) and (\ref{eq:straincross}).  This
regularizes the divergence of $\eta(t)$ at the coalescence time and
accounts for the fact that we may not have a clean separation between
the conventional amplitude and phase at the merger appearing in
Eqs.~(\ref{eq:strainplus}) and (\ref{eq:straincross}).  Since we are
allowed to perform appropriate diffeomorphisms of the control
parameters, we are led to reparametrizations of the phase shifted Airy
functions $\mathrm{Ai}_{\varphi_0}(\tau(t))$ and $\mathrm{Ai}_{\varphi_0-\pi/2}(\tau(t))$
defined above, respectively the analogues of the cosine and sine functions
in Eqs. (\ref{eq:strainplus}) and (\ref{eq:straincross}).
In this setting, we propose to modify Eqs.~(\ref{eq:strainplus}) and
(\ref{eq:straincross}) as
\begin{eqnarray}
  \label{eq:strainairy1}
  h_+(t) &=& A_+ a(t)\mathrm{Ai}_{\varphi_0}(\tau(t))\,,\\
    \label{eq:strainairy2}
  h_\times(t) &=& A_\times a(t)\mathrm{Ai}_{\varphi_0-\pi/2}(\tau(t))\,.  
\end{eqnarray}
This is the actual model we propose for the BBH waveform merger.  The
amplitude function $a(t)$ appearing here is distinct from $\eta(t)$
and remains finite at the merger.  For dimensional reasons, it must
still be proportional to $\frac{G\mathcal{M}}{c^2 D}$, but the
dependence on $t$ is modified.  This ansatz serves to describe a
general elliptically polarized gravitational wave.  In the
conventional model of Eqs.~(\ref{eq:strainplus}) and
(\ref{eq:straincross}), the binary parameters enter the phase
$\varphi(t)$.  Similarly, the system parameters enter the Airy
function model via the reparameterization function
$\tau(t)$~\footnote{It should be kept in mind that this corresponds to
  the transitional approximation.  We still have the flexibility to
  include the derivative of the Airy function (actually another
  appropriate diffraction integral $ J(\w{x})$ in
  Eq. (\ref{e:capital_J})) and a time dependent phase as given in
  Eq.~(\ref{e:Airy_gen}), at the price of including additional
  control parameters, e.g. the full spacetime coordinates $x^a$.}.
Note also that this model replaces the conventional complex signal
$\eta(t)e^{i\varphi_0}e^{i\varphi(t)}$ with
$e^{\pi i/3}e^{i\varphi_0}a(t)\mathrm{Ai}_+(t)$.  In regimes where the
amplitude changes much more slowly than the phase, both
representations are equivalent and in fact the conventional
representation is more convenient.  At the merger however the
situation is different and the Airy description becomes
necessary.

\subsubsection{Early Airy asymptotics and BBH inspiral}

At early times in the inspiral, it is not difficult to relate the
early asymptotics of the Airy function to the standard post-Newtonian
expressions.  To this end, compare the argument of the $\cos$ function
in Eq.~(\ref{eq:airyapprox1}) with the post-Newtonian phase given in
Eq.~(\ref{eq:pnphase}).  For
large $-t$ (so that both the $t_c$ and constant terms can be ignored),
we obtain
\begin{equation}
  (-\tau)^{3/2} = 3 \left(\frac{-t}{5G\mathcal{M}/c^3} \right)^{5/8}\,,
\end{equation}
which implies
\begin{equation}
  -\tau = 3^{2/3}\left(\frac{-t}{5G\mathcal{M}/c^3} \right)^{5/12}\,.  
\end{equation}
The amplitude is then also easily obtained
\begin{equation}
  a(t) \propto \left(\frac{-\tau(t)}{-t}\right)^{1/4} \propto (-t)^{-7/48}\,.
\end{equation}
We see that $a(t)\rightarrow 0$ as $t\rightarrow -\infty$.

\subsubsection{The merger time as the caustic time $\tau=0$}

Here we go beyond the post-Newtonian approximation and discuss the
merger itself.  It is conventional in the gravitational wave
literature to denote the peak of the waveform as the merger.  However,
if one accepts the point of view proposed here, this may be misleading.  The
peak of the Airy function occurs at (dimensionless) $\tau \approx -1.019$
and the merger, in the underlying caustic model,
a little later at the ``caustic time'' $\tau=0$. The latter is
the time at which the amplitude formally diverges in the `geometric optics'
description and when the transition from
oscillatory to damped behavior occurs.  We shall adopt this perspective and shall use the
terms ``merger time'' or ``caustic time'' for $\tau=0$~\footnote{It would also be
  consistent to refer to $\tau=0$ as the ``coalescence time'', since it corresponds
  to $t=t_c$ in Post-Newtonian expansions. We refer however to it as the ``merger time''
  in the present caustic model, since it is the only time naturally related to the
  merger in the present diffraction caustic framework.}.

\begin{figure}
\centering
\includegraphics[width=80mm,trim=0 6cm 0 6cm,clip]{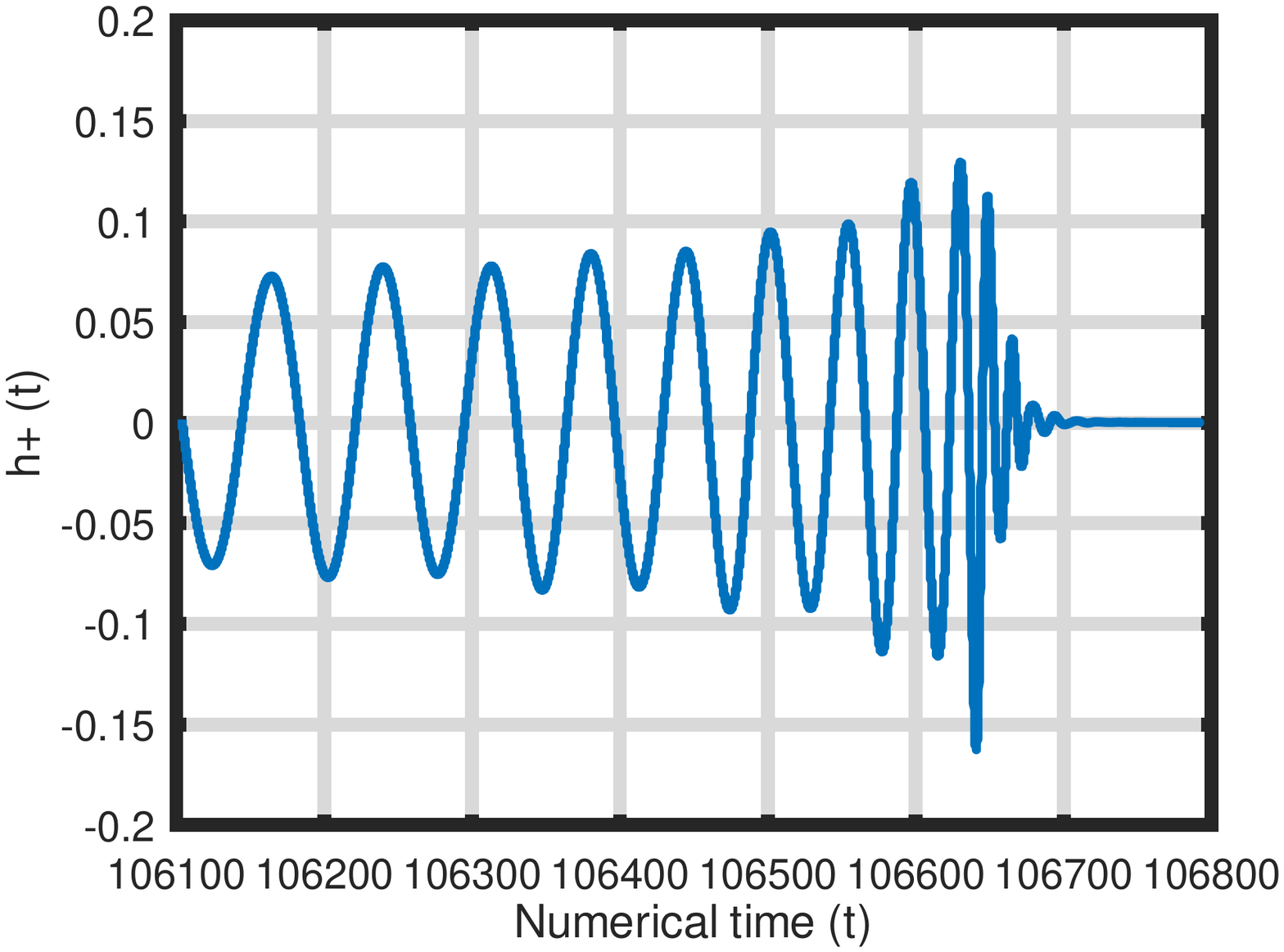}
\includegraphics[width=80mm,trim=0 6cm 0 6cm,clip]{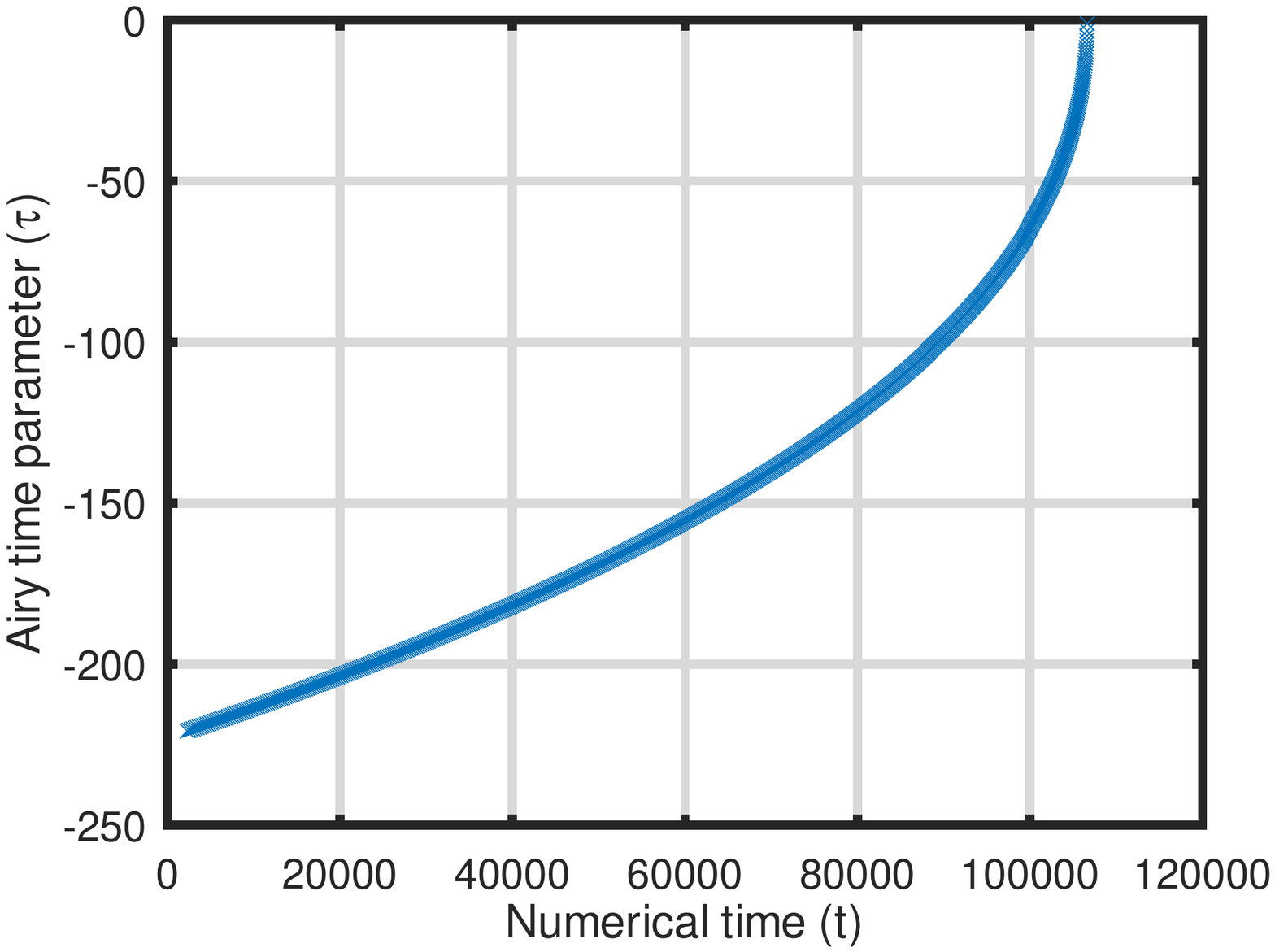}
\includegraphics[width=80mm,trim=0 6cm 0 6cm,clip]{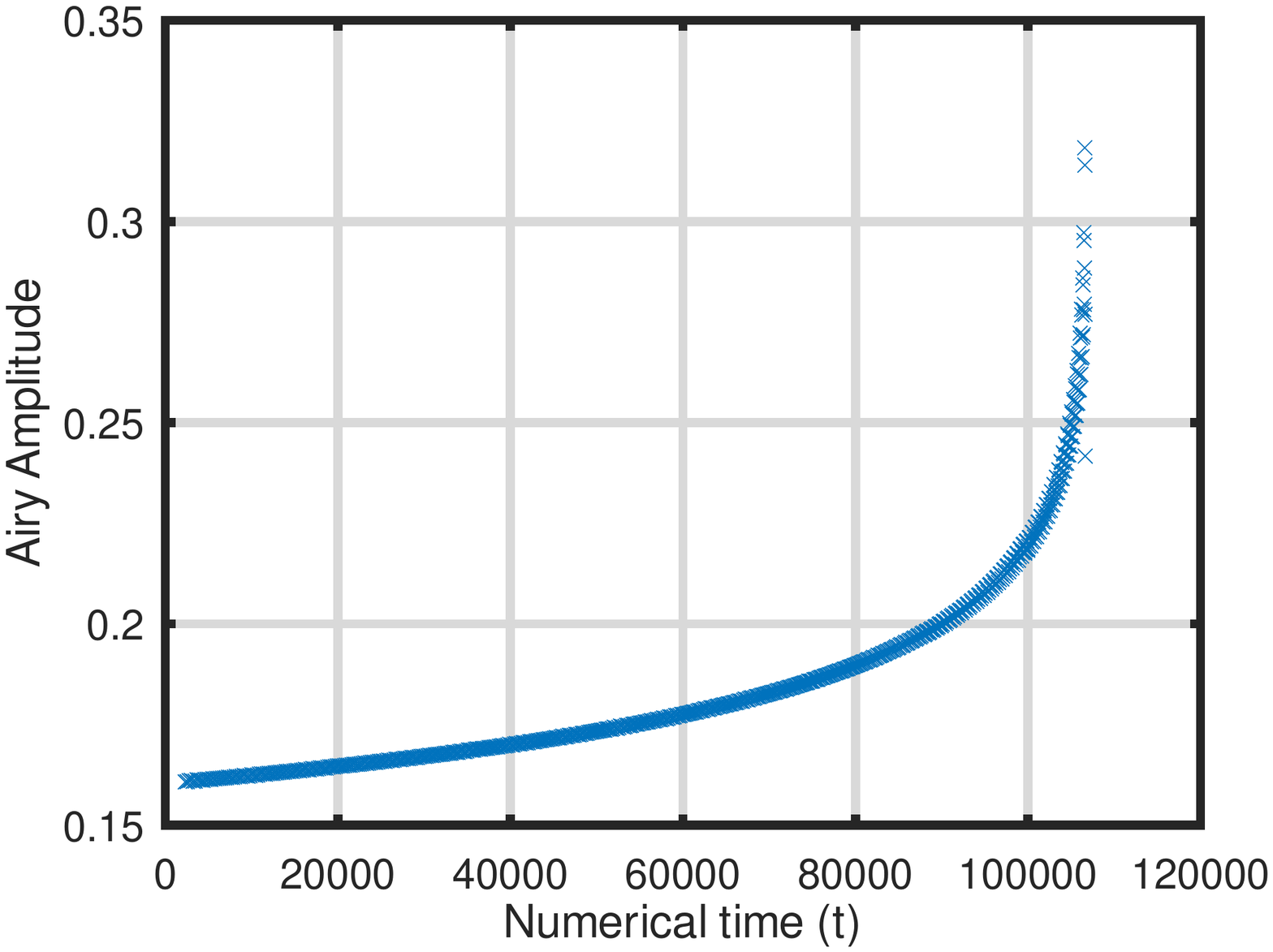}
\caption{The Airy reparametrization function and the Airy amplitude
  for a particular NR waveform from the SXS catalog. The first panel
  shows the late part of the waveform. The second panel is the time
  mapping function $\tau(t)$, and the last panel is the Airy amplitude
  $a(t)$ appearing in Eqs.~(\ref{eq:strainairy1}) and
  (\ref{eq:strainairy1}) following the procedure as described in the
  text.  Despite appearances, the values of $\tau$ in the second panel
  terminate at the Airy peak time $\tau=-1.019$, and do not reach
  $\tau=0$. }
\label{fig:mapping}
\end{figure}

Since the most accurate calculation of the merger signal to date is
via numerical simulations, we can compare the Airy function with a
numerical relativity signal.  We take a particular signal from the SXS
catalog \cite{Boyle:2019kee} corresponding to the merger of two
non-spinning black holes.  This is a ``hybrid'' waveform, constructed
by stitching together a numerical relativity waveform with the
corresponding post-Newtonian signal.  We align the peak of the
waveform with the largest peak of the Airy function.  Matching then
all preceding peaks and troughs of the Airy function for increasingly
negative times with the corresponding peaks and troughs of the $+$
polarization of the Numerical waveform leads to the mapping $\tau(t)$
shown in the second panel of Fig.~\ref{fig:mapping}.  Since the Airy function has no
oscillations or zero-crossings at later times after the peak, this
procedure cannot be extended to later times, and we must necessarily
resort to extrapolation.
We see that the function
$\tau(t)$ obtained from this procedure increases rapidly as
$\tau\rightarrow 0$.  This is a consequence of the fact that the
frequency of the Airy function decreases to zero as $\sqrt{\tau}$
while the frequency of the NR waveform increases monotonically.
Similarly, the amplitude also increases rapidly, as shown in
the third panel of Fig.~\ref{fig:mapping}.  Nevertheless, the
amplitude is a ratio of finite quantities, and it remains perfectly
finite.

A remarkable fact of curves $\tau(t)$ and $a(t)$ in, respectively,
the second and third panels of Fig.~\ref{fig:mapping}
is that the values of $\tau$ and $a$ remain finite at the caustic
time $t_c$ (corresponding to $\tau=0$), apparently reaching that
value with an infinite slope. This suggests that such
functions $\tau(t)$ and $a(t)$ are actually the lower branch 
of a multivalued function~\footnote{
  This raises the possibility of exploring the possibility of $\tau$ and/or $a$
  as appropriate ``state variables'' in the BBH fold caustic description,
  in the spirit of Fig. \ref{fig:Fold_geometry}.}. This would mean that the mapping
$t\mapsto \tau$ does not extend for $t>t_c$ (namely $\tau=0$), this entailing
the impossibility of realizing the Airy diffraction BBH model
beyond the caustic/merger point, consistently with our
initial set up of the problem, where
the ringdown must be accounted by a different physical mechanism
and suggests strongly that the ringdown mechanism and phase must be matched
at the merger/caustic time $\tau=0$.

\subsubsection{Transition to the ringdown at the caustic time}

Even though the Airy function model aims mainly at the merger phase,
the uniform approximation makes it also valuable in the inspiral
phase. On the contrary, as we have just mentioned, the late ringdown phase is, as a matter of
principle, not a part of this model.  It requires an entirely
different physical mechanism incorporating linear scattering
resonances (quasi-normal modes) of a finite size resonator (the final
merged black hole), in a linear ringdown phase.  Nevertheless,
assuming the validity of the Airy function model, we can still draw
some conclusions about the ringdown regime.

Note that the maximum of the caustic diffraction pattern,
  essentially given by the Airy function, occurs {\em before} the
caustic point at $\tau=0$. Soon after the caustic
point, we have seen that an Airy signal behaves as
$\sim e^{-\tau^{\frac{3}{2}}}$ (cf. Fig. \ref{fig:airyplot}), but this
over-exponential damping should be replaced by an
appropriate ringdown model.  This suggests a natural 
prediction of the model, namely an Airy-relation between the maximum
of the waveform and the transition to linearity signaled by the
``merger'' of the ``fold-caustic'' stationary points at $\tau=0$.  We are
led to the following conjecture for the transition from the merger
to the ringdown regime.  Several recent investigations have proposed
that the ringdown description of the post-merger signal in terms of
damped sinusoidal waveforms can be extended backwards in time towards
the merger, even up to the peak of the signal
\cite{Giesler:2019uxc,Isi:2022mhy} (see also
\cite{Cotesta:2022pci,Capano:2021etf}).  If the Airy function is
indeed an accurate representation of the signal, it suggests then the
possibility that the rapid decay of the Airy function after the peak
at $\tau \approx -1.019$ might mimic an exponentially damped signal
without oscillations.  The Airy function decreases monotonically from
$\approx 0.5357$ at its last peak to $\approx 0.3550$ at the caustic $\tau=0$.  This is
consistent with the fact that the oscillation frequencies found in
\cite{Giesler:2019uxc,Isi:2022mhy} are poorly constrained while the
damping time is found more accurately and is the dominant effect. On
the other hand, the true ringdown phase describing the final remnant
black hole is expected to become active only after the transition time
$\tau=0$, which occurs after the peak. The conjecture
is that this transition to the ringdown occurs exactly at
$1.019$ (dimensionless) units of Airy time $\tau$
after the peak, namely at the inflection point of the signal
when parametrised in the $\tau$ time.

\section{Conclusions}
\label{sec:conclusions}

Despite the non-linearities of general relativity, binary black hole
mergers waveforms are seen to be simple and to have universal
properties.  These include: i) the increase in amplitude and rapid
extinction shortly after the merger, ii) transition from an
oscillatory to a damped regime as we cross the merger, and iii) the
signal in the merger is seen to be relatively independent of the
initial conditions.  There exist separate approximation schemes in the
inspiral and ringdown regimes which are respectively the
post-Newtonian formalism and the framework of black hole perturbation
theory.  Both these formalisms break down as we go towards the merger.
In addition, there are further indications of surprising linearity in
the merger phase coming from numerical and observational studies of
the ringdown phase.

In this paper  the simplicity and universality of the BBH merger waveform
are addressed by adopting an approach that simplifies the details of the problem 
and focuses on its structurally stable aspects, in the spirit of an
`asymptotic reasoning'~\cite{Batte01,Batte97}. In particular, 
we explore the notion that the BBH merger waveform universal features are similar to
those found in numerous other natural phenomena, in particular in
optical phenomena involving caustics.
The framework of singularity
theory provides a classification of caustics under the assumption of
structural stability, and it also leads to useful expressions of the
radiation field in terms of universal functions given by Fresnel integrals
for diffraction over the caustics. Specifically, caustics in `geometric
optics' represent the ``geometrical skeleton supporting the wave
flesh''~\cite{KraOrl12},
the latter being provided by diffraction in `catastrophe optics'.  We have discussed the axisymmetric lens and the
rainbow as simple examples, and we have seen that the Airy function
associated with the fold catastrophe provides an appropriate model for
the BBH merger waveform. In particular, the associated Airy equation provides a
concrete realization of an effective linearity in this physical transient
process.  

We have argued that binary black hole merger waveforms share many of
the features of a fold caustic, and we are led to postulate that the
merger signal can be described as a reparameterized Airy function
$\mathrm{Ai}(\tau(t))$ and a slowly varying amplitude $a(t)$ which
remains finite at the merger.  At a practical level, our model is a
substitution of the conventional reparameterized sine/cosine functions
by the reparameterized Airy function.  We have proposed a concrete
model for the binary merger waveforms given in
Eqs.~(\ref{eq:strainairy1}) and (\ref{eq:strainairy2}) meant to
capture the inspiral-merger transition.  This is not (yet) a full
fledged inspiral-merger-ringdown model optimized for use in
gravitational wave data analysis.

The Airy function has a natural time $\tau=0$ which can be identified
as the merger time which separates the oscillatory and damped regimes.
However, the peak of the Airy function occurs \emph{before} $\tau=0$.
Thus, we conjecture that the common identification of the waveform
peak with the merger time is incorrect.  From this perspective, it is
natural to attach an appropriate ringdown model only for $\tau>0$.
This has important consequences, for example, for issues related with
black hole spectroscopy and the identification of ringdown overtones.
Just like the phase function in the inspiral regime, the
reparameterization function $\tau(t)$ and the amplitude $a(t)$ depend
on the binary parameters.  A first comparison of $\tau(t)$ and $a(t)$
with the leading order post-Newtonian result and with numerically
generated merger waveforms can be carried out. However, much more
remains to be done; the detailed dependence of $(\tau(t),a(t))$ on the
mass ratio, spins, eccentricity will be explored elsewhere, and this
will be critical for applications to gravitational wave astronomy and
data analysis.

It is important to keep in mind that the ringdown itself is a distinct
phenomena and needs to be added separately; it depends on the
finite-size nature of the remnant object and is not automatically part
of the present catastrophe theory framework built on the basis of rays
in geometric optics.  Thus, as a first approximation, neutron star
mergers share several of these features for the inspiral and merger
regimes.  However, unlike binary black hole mergers, the remnant is
very different.  The phenomena which model this regime are clearly
very different from what we see in a binary black hole merger remnant.
Nevertheless, the Airy function model could be used to link an
appropriate post-merger model with the inspiral.  

We have argued that similar behavior should be viable in alternate
theories of gravity as well.  Catastrophe theory contains the notion
of structural stability into its fundamental assumptions which means
that small perturbations should not lead to qualitatively different
behavior.  In fact, it can be argued that none of our models of a
physical system can be \emph{exactly} true and all of our models
should satisfy the notion of structural stability in order to be
viable.  This leads us to conjecture that binary black hole merger
waveforms in alternate theories of gravity should share the features
of simplicity and universality seen in standard general relativity.
In fact, we can include deviations from standard general relativity:
the parameterized post-Newtonian framework (including diffeomorphism
invariant metric theories of gravity) results in qualitatively similar
behavior for the pre-merger regime.  Similarly, it is reasonable to
postulate that the remnant black hole in many alternate theories of
gravity should approach equilibrium by emitting damped sinusoidal
radiation.  We are therefore again led to the transition from an
oscillatory to damped behavior, and catastrophe theory again leads to
the Airy function model.

While the immediate goals are to develop and test this model with a
view towards applications in gravitational wave astronomy, it is
plausible that this work will enable us to discover new physical
phenomena connected with the merger.  One of these is the Arnold
exponent $\beta$ which appears in the scaling laws of the amplitude at
the merger.  From the Airy function model, we predict that
$\beta=1/6$.  An independent confirmation via numerics and/or analytic
work (and eventually by observations) would be of interest. More
broadly, the ultimate goal is to derive this model from first
principles starting with the Einstein equations and following
systematic approximation schemes.  In this setting it is tantalizing
to consider a specially important non-linear generalization of the Airy
function, namely the Painlev\'e-II Transcendent, intimately associated
to integrable systems \cite{ConMus08}. In particular, it will be of
interest to investigate the role of integrability in binary black hole
dynamics and its implications for universality \cite{JarKri22bv2}.

\section*{Acknowledgments}

We thank Abhay Ashtekar, Ivan Booth and Andrey Shoom for discussions
at the early stage of this project, at the ``Focus Session: Dynamical
Horizons, Binary Coalescences, Simulations and Waveform'' at the
Pennsylvania State University in July 2018. We are especially indebted to Oscar
Meneses-Rojas and Ricardo Uribe-Vargas for the patient and generous
sharing of their insights in singularity theory.  We also thank Bruce
Allen, Carlos Barcel\'o, Beatrice Bonga, Sam Dolan, Vincent Lam, Klaas
Landsman, Peter Miller, Alex Nielsen, Ariadna Ribes-Metideri,
Mikhail Semenov Tian-Shansky, Dhruv
Sharma, Carlos F. Sopuerta and Th\'eo
Torres for valuable discussions.  We also thank the Banff
International Research Station for enabling numerous discussions
during the Workshop ``At the Interface of Mathematical Relativity and
Astrophysics''.  We are grateful to all participants in this workshop.
This work was supported by the French ``Investissements d’Avenir''
program through project ISITE-BFC (ANR-15-IDEX-03), the ANR ``Quantum
Fields interacting with Geometry'' (QFG) project
(ANR-20-CE40-0018-02), the EIPHI Graduate School (ANR-17- EURE-0002)
and the Spanish FIS2017-86497-C2-1 project (with FEDER contribution).


\medskip


\bibliography{fAiry_Fold_blue}




\end{document}